\documentclass[twocolumn, trackchanges]{aastex701}

\usepackage{float}

\begin{document}

\title{MAGAZ3NE: Spatially Resolved Ages and Chemical Abundances of Ultra-Massive Quiescent Galaxies at z $\sim$ 3.5 using JWST/NIRSpec IFU}

\correspondingauthor{Adit H. Edward}

\author[0009-0000-3748-9792]{Adit H. Edward}
\affiliation{Department of Physics and Astronomy, York University,
4700 Keele St. Toronto, Ontario, M3J 1P3, Canada}
\email[show]{ahedward@yorku.ca}  

\author[0000-0002-0243-6575]{Jacqueline Antwi-Danso} 
\altaffiliation{Dunlap Fellow}
\affiliation{David A. Dunlap Department of Astronomy and Astrophysics, University of Toronto, 50 St. George Street, Toronto, Ontario, M5S 3H4, Canada}
\affiliation{Dunlap Institute for Astronomy \& Astrophysics, University of Toronto, 50 St George Street, Toronto, ON M5S 3H4, Canada}
\affiliation{Department of Astronomy, University of Massachusetts, Amherst, MA 01003, USA}
\email{j.antwidanso@utoronto.ca}  

\author[0000-0002-9330-9108]{Adam Muzzin}
\affiliation{Department of Physics and Astronomy, York University, 
4700 Keele St. Toronto, Ontario, M3J 1P3, Canada}
\email{muzzin@yorku.ca}

\author[0000-0001-6003-0541]{Ben Forrest}
\affiliation{Department of Physics and Astronomy, University of California Davis, One Shields Avenue, Davis, CA, 95616, USA}
\email{bforrest@ucdavis.edu}

\author[0000-0002-2446-8770]{Ian McConachie}
\affiliation{Department of Astronomy, University of Wisconsin-Madison, 475 N. Charter St., Madison, WI 53706 USA}
\email{ian.mcconachie@wisc.edu}

\author[0000-0002-9861-4515]{Aliza Beverage}
\altaffiliation{NHFP Hubble Fellow}
\affiliation{Observatories of the Carnegie Institution for Science, 813 Santa Barbara Street, Pasadena, CA 91101, USA}
\affiliation{Department of Astrophysical Sciences, Princeton University, 4 Ivy Lane, Princeton, NJ 08544, USA}
\email{ abeverage@carnegiescience.edu}

\author[0000-0003-2144-2943]{Wenjun Chang}
\affiliation{Department of Physics and Astronomy, University of California, Riverside, 900 University Avenue, Riverside, CA 92521, USA}
\email{wchan148@ucr.edu}

\author[0000-0003-1371-6019]{M. C. Cooper}
\affiliation{Center for Cosmology, Department of Physics and Astronomy, University of California, Irvine, Irvine, CA, USA}
\email{cooper@uci.edu}

\author[]{Percy Gomez}
\affiliation{W.M. Keck Observatory, 65-1120 Mamalahoa Hwy., Kamuela, HI 96743, USA}
\email{pgomez@keck.hawaii.edu}

\author[0000-0001-6763-5551]{Massissilia L. Hamadouche}
\affiliation{Department of Astronomy, University of Massachusetts, Amherst, MA 01003, USA}
\email{mhamadouche@umass.edu}

\author[0000-0002-9466-2763]{Aur{\'e}lien Henry}
\affiliation{Department of Physics, University of California, Merced, 5200 Lake Road, Merced, CA 95343, USA}
\email{ahenry13@ucmerced.edu}

\author[0000-0002-9158-6996]{Han Lei}
\affiliation{Department of Physics, McGill Space Institute, McGill University, 3600 rue University, Montr\'eal, Qu\'ebec H3A 2T8, Canada}
\email{han.lei@mail.mcgill.ca}

\author[0000-0001-9002-3502]{Danilo Marchesini}
\affiliation{Department of Physics \& Astronomy, Tufts University, MA 02155, USA}
\email{Danilo.Marchesini@tufts.edu}

\author[]{Allison Noble}
\affiliation{School of Earth and Space Exploration, Arizona State University, Tempe, AZ 85287, USA}
\affiliation{Beus Center for Cosmic Foundations, Arizona State University, Tempe, AZ 85287 USA}
\email{anoble5@asu.edu}

\author[]{Stephanie M. Urbano Stawinski}
\affiliation{Department of Physics, University of California, Santa Barbara, Santa Barbara, CA 93106, USA}
\email{sstawins@ucsb.edu}

\author[0000-0002-6572-7089]{Gillian Wilson}
\affiliation{Department of Physics, University of California, Merced, 5200 Lake Road, Merced, CA 95343, USA}
\email{gwilson@ucmerced.edu}

\author[0000-0002-6505-9981]{M. E. Wisz}
\affiliation{Department of Physics, University of California, Merced, 5200 Lake Road, Merced, CA 95343, USA}
\email{mwisz@ucmerced.edu}


\begin{abstract}

We present spatially-resolved measurements of stellar age, [Fe/H], and [$\alpha$/Fe] in three ultra-massive ($\rm{log(M_{\ast}/M_{\odot})>11}$), compact ($\rm{R_e} \lesssim 2$ kpc) quiescent galaxies at $z\sim3.5$ using JWST/NIRSpec IFU spectroscopy. These observations provide the first spatially-resolved constraints on $\alpha$-enhancement at this epoch, enabling a direct test of quenching mechanisms before late-time assembly processes such as mergers can erase chemical signatures. The central regions of all three galaxies show both uniformly young ages ($\approx0.6-0.7$ Gyr) and elevated [$\alpha$/Fe] ($\approx0.2-0.5$), indicating rapid, enhanced star formation shortly before recent quenching. Beyond the cores, two galaxies display positive age gradients and negative [$\alpha$/Fe] gradients, consistent with rapid merger-driven quenching, while the third shows a flat age profile indicative of uniform quenching. The [Fe/H] gradients are also consistent with these trends, though we note that the metallicities reported by codes using $\alpha$-enhanced models differ significantly ($\approx0.2-0.4$ dex) from those reported using solar-scaled templates.
These data demonstrate that quenching pathways are diverse by $z\sim3.5$, with rapid, merger-driven quenching already operating in a subset of massive quiescent galaxies in the first two billion years of cosmic time. 
Furthermore, these results establish that explicit treatment of $\alpha$-enhancement is essential for interpreting the star-formation histories of the earliest quenched systems.

\end{abstract}



\section{Introduction} \label{sec:intro}


Since the first discoveries of massive quiescent galaxies (MQGs) in the early Universe \citep[e.g.,][]{Dunlop96, Glazebrook04, Daddi05, Kriek06, Kriek08, Kriek09}, they have remained one of the most active areas of research in the following decades \citep[e.g.,][]{vanDokkum08, Bezanson09, Marchesini10, Newman10, Brammer11, Man12, Muzzin12, vandeSande13, Whitaker13, Straatman14, Belli14, Belli15, Newman15, Kriek16, Glazebrook17, Marsan17, Belli17, Newman18, Kriek19, Belli19, Kriek19, Akhshik20, Jafariyazani20, Forrest20a, Forrest22, Akhshik23, UrbanoStawinski24, Ito25, Antwi-Danso25, Jafariyazani25}. 

The James Webb Space Telescope (JWST) has already advanced this field significantly, identifying MQGs at earlier times than was possible with ground-based telescopes \citep[e.g.,][]{ Carnall23MQG, Carnall23, Carnall24, Weibel24, Nanayakkara24, DEugenio24, Wright24, Baker25b, deGraff25, Yang26, Hamadouche26, Ji26, Zhang26b}.
In particular, high signal-to-noise JWST spectroscopy has enabled more precise star formation history (SFH) modeling, improving insights into past star-formation and quenching \citep[e.g.,][]{Park24, Glazebrook24, Beverage24, Beverage25, Cheng25, McConachie25, Hamadouche26, Lisiecki26, Zhang26a}. 
However, JWST observations of early MQGs have heightened previous tensions between observations and theoretical galaxy formation models for these extreme objects. 
Their inferred SFHs indicate earlier and more rapid formation than what is expected from both semi-analytic models and hydrodynamical simulations \citep[i.e.,][]{Schreiber18, Glazebrook24, Nanayakkara24, deGraff25}. Resolving this tension requires improved constraints on their formation timescales, as well as a clearer understanding of the physical mechanisms that quenched them at such early times.

In contrast to high-redshift work, the quenching histories of massive quiescent galaxies in the local Universe have been well-characterized. From these works, we have identified three broad quenching pathways: uniform, outside-in, and inside-out. 
In uniform quenching,  the entire galaxy is quenched by the same mechanism on very short timescales of $\lesssim 100$ Myr \citep{Setton20, DEugenio24}, usually presumed from active galactic nuclei (AGN) feedback.
Post-starburst (PSB) galaxies at $z\sim0.6$ have shown this to be a dominant quenching mechanism \citep{Setton20}.
In the outside-in scenario, the galaxy quenches its star formation in the outskirts before the cores, typically from environmental effects such as tidal interactions \citep[e.g.,][]{Schaefer17}. This leaves the stellar populations on the outskirts older than those in the central regions. 
Finally, inside-out quenching describes the process whereby galaxies quench their inner regions first, leaving the stellar populations in the core older than those on the periphery. 
On average, the cores of galaxies build up their mass earlier than their outskirts \citep{Garcia17}, independent of Hubble type, surface density, and stellar mass. 
Many local ellipticals are predicted to have formed through this process, due to the numerous mergers and accretions they experienced in the past \citep[e.g,][]{Nelson16, Oldham17, Zewdie21, McDonough25, Pizzardo25, Lawlor-Forsyth26}.

Simulations provide a fourth pathway for high-$z$ galaxies: rapid, merger-driven quenching \citep{Pathak21, Kimmig25,  Ni25}. In this scenario, galaxies undergo a compaction event where the gas funnels into the center of the galaxy, triggering an intense central starburst. As the gas reservoir depletes, AGN feedback suppresses further star formation, quenching the galaxy in an inside-out manner. This process occurs rapidly, quenching on timescales of $\lesssim200$ Myr, often following a major merger. 
Despite the inside-out sequence of quenching, it produces younger light-weighted cores due to the recent starburst \citep{Pathak21, Ni25}.

Discerning between these different quenching pathways at high redshift requires spatially resolved measurements of age, metallicity, and $\alpha$-enhancement. In particular, $\alpha$-enhancement is an important tracer of formation histories. The $\alpha$ elements (i.e., C, O, Ne, Mg, Si, S, Ar, Ca, Ti) are typically produced in massive stars before being released by core-collapse supernovae (SNe II) that enrich the interstellar medium (ISM) on short \citep[$\lesssim$ 30 Myr;][]{Pipino09} timescales. This is in contrast to Fe-peak elements, which are produced in equal amounts by both SNe II and Type Ia supernovae (SNe Ia), although the latter occurs on much longer timescales of $\sim0.5-1$ Gyr \citep{Maoz04, Maoz10, Kobayashi09}. It follows then that the abundance ratio [$\alpha$/Fe]\footnote{[A/B] = log$(n(A)/n(B))_{\ast} -$ log$(n(A)/n(B))_{\odot}$, where $n(X)$ is the number density of element X.} 
is a \emph{direct} tracer of the formation timescales of young ($<1$ Gyr old) galaxies. 
\citet{Beverage25} for instance show that star-formation timescales inferred with variable abundances are shorter than those that assume solar-scaled ([$\alpha$/Fe]$=0$) abundances.

The combination of age, metallicity, and [$\alpha$/Fe] gradients encodes information on the various quenching pathways for quiescent galaxies \citep{M+M19}. 
Measuring these gradients was difficult in the pre-JWST era, due to the high spatial and spectral resolution required, necessitating long exposure times. 
Deep spectroscopic surveys were therefore essential to characterize these trends statistically. 
\citet{Cheng24} presented age, [Fe/H], and [Mg/Fe] radial gradients for 456 massive quiescent galaxies at $0.6 \lesssim z \lesssim 1.0 $ from the Large Early Galaxy Astrophysics Census \citep[LEGA-C;][]{vdWel16, vdWel21, Straatman18}, finding flat age and [Mg/Fe] gradients, as well as negative [Fe/H] gradients. They also found that younger ($\rm 1  \leq  age/Gyr < 2.63$ ) quiescent galaxies have negative [Fe/H] and positive age gradients, consistent with a central merger-driven starburst, although minor mergers and progenitor bias \citep{vDokk01, Belli15, Keating15} can produce similar signatures. 
Follow-up work from the JWST-Spectroscopic Ultradeep Survey Probing Extragalactic Near-infrared Stellar Emission \citep[SUSPENSE;][]{Slob24} studied eight massive ($10.3 \lesssim \mathrm{log}(M_{\ast}/M_{\odot}) \lesssim 11.1$) quiescent galaxies at $1.2 \lesssim z \lesssim 2.2$ \citep{Cheng25}. They found negative age gradients, positive [Mg/Fe] gradients, and flat [Fe/H] gradients, pointing to inside-out quenching.

In contrast, very few constraints on quenching pathways exist at higher redshifts ($z \gtrsim 2$). The few studies that have measured resolved stellar population parameters reveal a diverse picture. For example, \citet{P-G25} found negative [Fe/H] gradients, along with evidence of inside-out formation and high [Mg/Fe] abundances in Jekyll, a well-studied massive  quiescent galaxy at $z\sim3.7$ \citep{Schreiber18, Glazebrook24, Nanayakkara24}. \citet{Takahashi25}, on the other hand, found that the line-emitting region (a tracer of on-going star-formation) for a ``mini-quenched" \citep[e.g.,][]{Looser24} galaxy at $z\sim5$ was more compact than the continuum region (a tracer of stellar mass), implying outside-in quenching. \citet{DEugenio24} found flat age evolution in a young ($\sim0.5$ Gyr), massive ($\mathrm{log(M_{\ast}/M_{\odot})} \approx 11.2$), PSB at $z\sim3$, which they show is most likely due to AGN feedback from the central supermassive black hole.
Understanding the relative role of these different quenching processes requires deep, spatially resolved observations of their stellar light at early cosmic times, prior to late-time assembly. 

To this end, we present the first spatially resolved measurements of [$\alpha$/Fe] for three massive ($\rm log(M_{\ast} / M_{\odot}) > 11$) quiescent galaxies at $z > 3$ using JWST NIRSpec Integral Field Unit (IFU) medium grating (R $\sim 1000$) observations.  With its high spatial resolution (spaxel size of 0.1$^{\prime\prime}$; \citealt{Boker22}), the IFU also enables age, [Fe/H], and SFH gradients. We can therefore obtain a more complete picture of the rapid evolution of these extreme objects. These observations allow us to directly test quenching pathways in the first 2 Gyr. The paper is structured as follows. In \S \ref{sec:data}, we present the observations. In \S \ref{sec:methods} and  \S \ref{sec:results}, we discuss our SED fitting methods and results. Finally, in \S \ref{sec:discuss} we discuss the implications for massive galaxy formation at high redshift. We summarize our main conclusions in \S \ref{sec:conc}. Throughout, we assume a \citet{KroupaBoily02} Initial Mass Function (IMF) and assume a flat $\Lambda$CDM cosmology, i.e., $H_{0} = 70$ km s$^{-1}$ Mpc$^{-1}$, $\Omega_{M} = 0.3$, and $\Omega_{\Lambda} = 0.7$.

\section{Data} \label{sec:data}

\subsection{JWST GO 2913} \label{sec:JWSTGO2913}


Our targets are three bright ($K_{s} < 21.7$) massive (log($M_{\ast}/M_{\odot}$) $> 11$) quiescent galaxies (MQGs) from the Massive Ancient Galaxies At $z>3$ NEar-infrared (MAGAZ3NE) Survey \citep{Forrest20, Forrest24}. MAGAZ3NE targets were photometrically identified from the UltraVISTA \citep{Muzzin13} and XMM-VIDEO (Annunziatella et al. in prep) catalogs and spectroscopically confirmed with Keck/MOSFIRE. 
Further spectroscopic follow-up with both Keck/MOSFIRE and Keck/NIRES 
revealed large ($\sigma_{\ast} \gtrsim 379$ km/s) velocity dispersions \citep{Forrest22}. 
All three MQGs have subsequently had deep follow-up ALMA Band 7 observations targeting the dust-continuum \citep{Chang26}. Two of the three were undetected, and only a faint detection was observed for the third (XMM-VID3-2457), confirming their quiescent nature against the possibility of dust obscured star-formation.

In this work, we utilize data from a JWST Cycle 2 program \citep[ID: GO 2913; `Dissecting the Monsters: Resolved IFU Spectroscopy of the Most Massive Quiescent Galaxies at $z>3$',][]{Forrest23}. This program obtained JWST/NIRSpec Integral Field Unit (IFU) observations using the G235M/F170LP grating-filter configuration, which covers $\rm 1.66 < \lambda_{obs} /{\mu}m < 3.17$. 
At  $3.45<z<3.5$, this corresponds to $\rm \sim 0.37 < \lambda_{rest}/\mu m < 0.70$, capturing important rest-frame optical features sensitive to age (e.g., the Balmer series \& $D_n$4000), metallicity (e.g., $\mathrm{Fe}5270$ \& $\mathrm{Fe}5335$) and $\alpha$-enhancement (e.g., Mg$b$, Ca I) \citep{Hamadouche26b}. 
The IFU spaxel size of 0.1$^{\prime\prime}$ is equivalent to $\approx730$pc at $z\sim3.5$ \citep{Forrest25}, enabling spatially resolved analyses of these MQGs.


We use the same data reduction as \citet{Forrest25}, which first presented these data. 
To summarize, data reduction was initially performed with the JWST pipeline \citep{Bushouse_JWST}. To mask contaminating flux, cosmic rays and hot pixels, we flag pixels that are $>2\sigma_{\mathrm{NMAD}}$  above the median flux for both the image and the spectrum for that spaxel, as well as the presence of an adjacent pixel flagged by the pipeline, where $\sigma_{\rm NMAD}$ is defined as the normalized median absolute deviation.

\begin{figure*}[!bthp]
\includegraphics[width=\linewidth]{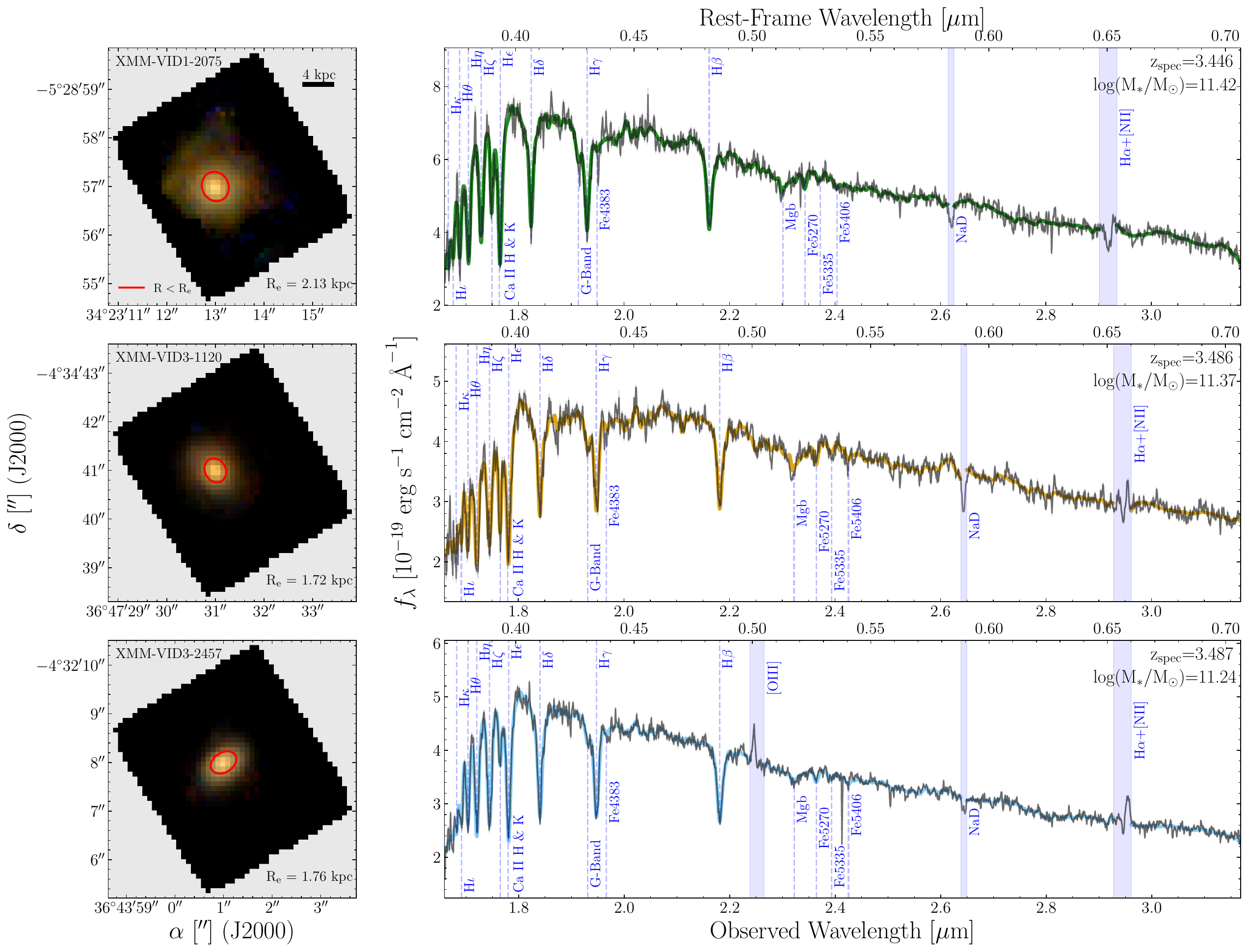} 
\centering
\cprotect\caption{\textit{Left}: False-color images of the MAGAZ3NE massive quiescent galaxies. We collapse the IFU data cubes along the wavelength axis and take the mean value for each pixel in the wavelength ranges of (6250\AA, 7500\AA), (4000\AA, 5600\AA) and (3735\AA, 4000\AA) for the R, G and B channels, respectively. We outline the effective radius (R$\mathrm{_{e}}$) of each galaxy in red and list the circularized R$_{\rm e}$ in kpc from \cite{Forrest25} in the bottom right. All three galaxies are very compact, with $\rm R_e \lesssim 2$ kpc. Of particular note is XMM-VID1-2075, whose image shows low-surface brightness structure, indicative of a recent major merger.
\textit{Right}: Integrated ($\mathrm{R < 1 R_{e}}$) 1D spectra with select spectral features highlighted. The associated uncertainty is shown in gray. Regions masked during SED fitting are shaded. The best-fit \verb|alf|$\alpha$ models are plotted as well (XMM-VID1-2075, \textit{green}; XMM-VID3-1120, \textit{gold}; XMM-VID3-2457, \textit{blue}). The galaxies all exhibit strong Balmer breaks and deep Balmer absorption lines, indicative of recently-quenched stellar populations. 
}
\label{fig:RGB}
\end{figure*}

We present false-color images of the IFU data cubes collapsed along the wavelength axis in Figure \ref{fig:RGB}, along with the extracted 1D spectra. We create these images by taking the mean value for the flux of each pixel in the wavelength ranges of (6250\AA, 7500\AA), (4000\AA, 5600\AA) and (3735\AA, 4000\AA) for the red, green and blue channels, respectively. 
The integrated 1D spectra are extracted by coadding the flux from all the spaxels within an ellipse corresponding to the circularized half-light radius ($\rm R_e$). 
We additionally coadd spaxels in elliptical annuli with widths of $\frac{1}{3}$ of the semi-major axis ($\rm R_{e,maj}$), equivalent to $\sim1$ pixel for each galaxy.
This is done for 5 bins for each MQG, with the furthest annuli extending to $\frac{5}{3}\rm R_{e,maj}$.
Integrated 1D spectra for each MQG are shown in Appendix \ref{sec:App_spec}. 

\begin{figure}[!htb]
\includegraphics[width=\columnwidth]{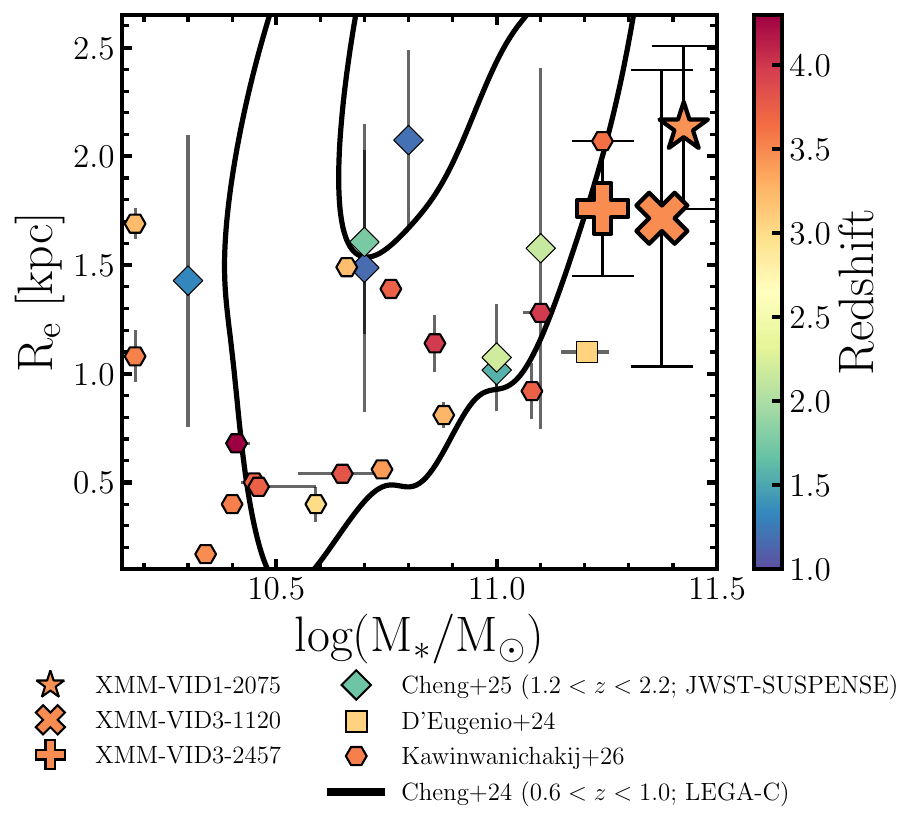}
\centering
\cprotect\caption{Size vs. mass comparison of massive quiescent galaxies both from this work and in literature, colored by redshift. Galaxies used in \citet{Cheng25} from JWST-SUSPENSE \citep{Slob24} are shown as diamonds, and the galaxy studied in \citet{DEugenio24} is shown as a square, with reported error bars. Galaxies from \citet{Kawinwanichakij26} are shown as hexagons, using the inferred $\rm R_e$ from F277W JWST observations. 
Contours show the 1 and 2$\sigma$ contours for the LEGA-C \citep{vdWel21} massive quiescent galaxy sample from \citet{Cheng24}, with masses taken from \citet{deGraff21}. 
}
\label{fig:mass-size}
\end{figure}

These galaxies are among the most massive ($11.23 \lesssim \mathrm{log(M_{\ast}/M_{\odot})} \lesssim 11.50$) of the MAGAZ3NE survey, which targeted 32 galaxies. 
To highlight this, we present the size versus mass plot for the three MQGs in this study, along with those from similar studies in literature in Figure \ref{fig:mass-size}, colored by redshift. The galaxies in this study are both more massive and more compact than those at lower-$z$. This informs us that the galaxies in our sample are not evolutionarily linked to those shown in lower-$z$ studies. This is discussed further in Section \ref{sec:lowerz_lit}.

\section{Methods} \label{sec:methods}

\subsection{Bagpipes SED Fitting} \label{Sec: Bagpipes}

We use the Bayesian Analysis of Galaxies for Physical Inference and Parameter EStimation \citep[\textsc{Bagpipes}\footnote{https://github.com/ACCarnall/bagpipes};][]{Carnall18, Carnall19} to perform spectral fitting. \textsc{Bagpipes} models complex galaxy properties and star formation histories within a Bayesian framework \citep{Carnall18, Carnall19}. We use the default stellar library, the 2016 version of the \citet{BC03} stellar population models. We note that this stellar library assumes solar-scaled abundances. 

We adopt a flexible non-parametric SFH with N=10 time bins to capture potential early bursts using the \texttt{Dense Basis} \citep{Iyer19} implementation in \textsc{Bagpipes}. This method reconstructs galaxy star formation histories using Gaussian Processes. This Gaussian Process SFH (GP-SFH) is parameterized as the stellar mass, star formation rate (SFR), and N lookback times ($\{ t_{x} \}$) corresponding to when the galaxy formed stellar mass fractions (i.e., $t_{10}$ corresponds to the lookback time when 10\% of the total stellar mass is formed, $t_{20}$ corresponds to the lookback time at which 20\% was formed, etc). These lookback times have a Dirichlet prior \citep{Iyer19}, and a corresponding $\alpha$ concentration parameter. When $\alpha < 1$, lookback times are allowed to be granular, resulting in `bursty' SFHs. When $\alpha > 1$, the lookback times are more evenly spaced out, which results in smoother SFHs \citep{Iyer19}. Similar to \citet{Hamadouche26}, we set $\alpha=30$ in the most recent time bin, and $\alpha=3$ in the remaining 9. This is to avoid the `first-bin burstiness' that can affect the \textsc{Bagpipes} implementation of \texttt{Dense Basis} (see \citealt{McConachie25}).  

We assume a \citet{Calzetti00} dust attenuation law, and an AGN component for XMM-VID3-2457 due to the presence of \textsc{[O iii]}$\lambda\lambda$4959,5007 (EW$_{\rm rest}$ = 3.6 \AA) and \textsc{[N ii]}$\lambda\lambda6548,6584$ (EW$_{\rm rest}$ = 3.5 \AA) emission lines. Instrumental dispersion is accounted for by incorporating the wavelength-dependent resolution curve for JWST/NIRSpec IFU G235M/F170LP into the fit. The priors are summarized in Appendix \ref{sec:priors}. In short, we vary redshift, velocity dispersion, stellar mass, SFR, metallicity, dust attenuation, and the nebular ionization parameter, $\rm log_{10}(U)$. We use the same Gaussian Process (GP) noise model first presented in \citep{Leung24}. This model uses \texttt{celerite2} \citep{celerite1, celerite2} and its SHOTerm kernel for correlated and white noise. This GP noise model is treated as an additive term to the spectrum to account for observational uncertainties and mismatched stellar spectral models.

To ensure that our results are robust to the choice of SFH prior (see \citealt{Webb22}), we additionally adopt the continuity model \citep{Leja19} integrated into \textsc{Bagpipes}. This model allows the user to set their own N time bins and their widths, and fits directly for the difference in SFR between adjoining bins. The continuity model does this by assuming a student-$t$ distribution on the quantity $x = \mathrm{log(SFR_{n}/SFR_{n+1})}$, where n refers to a given time bin \citep{Leja19}. The GP-SFH model is  better at capturing bursty or episodic star formation histories \citep{Iyer19}, while the continuity model prefers smoother histories unless required by the data. As both models encode different star formation history parameterizations, agreement between them would suggest that our results are data-driven rather than prior-dominated.

\citet{Leja19} showed that when $\rm N \gtrsim 4$, SFH results are insensitive to this choice. Similarly, \citet{Iyer19} showed that $\rm N = 3$ is sufficient to recover SFHs, but the accuracy increases with N. With the high quality of spectra available to us, we are able to test the sensitivity of our SFH constraints to the adopted non-parametric parameterization. For this reason, we use 10 time bins in our implementation of the continuity SFH model. The first 3 bins are spaced out linearly by 50 Myr, while the remaining bins are spaced out equally in log-space to the time of observation. We fix the redshift to the posterior median value from the GP-SFH fit for each MQG. We adopt a scale factor of $\sigma=0.7$ to balance between bursty and smooth SFHs. We also set the degrees of freedom $\nu=2$ as in \citet{Leja19}. Elsewhere, we fit for stellar population parameters in a similar manner as the GP-SFH fit (see Appendix \ref{sec:priors}).

\subsection{\texorpdfstring{alf$\alpha$}{alfa}} \label{sec:AA}

The ultimate goal of SED fitting is to infer stellar population parameters and SFHs, but the age-dust-metallicity degeneracy limits any interpretation that is unable to break it. As most SED fitting codes use solar-scaled models, it is difficult to accurately constrain metallicities. Solar-scaled models do not capture variations in different elemental species, such as $\alpha$-elements. However, chemical abundance codes address this by allowing elemental abundances to vary, and therefore better disentangle the metallicity from degeneracies with other elements.

We use \verb|alf|$\alpha$\footnote{https://github.com/alizabeverage/alfalpha} \citep{Beverage2024_AA, Beverage25} to derive ages, [$\alpha$/Fe] \& metallicity. \verb|alf|$\alpha$ is a Python version of the full-spectrum fitting code Absorption Line Fitter \citep[\texttt{ALF};][]{Conroy12, Conroy18}, and incorporates the same single stellar population (SSP) models and elemental response functions presented in \citet{Conroy18}. The SSPs are built from both the empirical MILES \citep{MILES} and extended IRTF \citep{Villaume17} stellar libraries, and utilize the MIST v1.2 \citep{MISTv1_2} isochrones. Additionally, to obtain elemental abundances, \citet{Conroy18} provided response functions for 19 elements. \verb|alf|$\alpha$ models the SFH as a single burst, effectively returning an SSP-equivalent age.

However, the \citet{Conroy18} models are restricted to stellar populations that are older than 1 Gyr, and so are not appropriate for fitting the young quiescent galaxies observed at high redshift \citep[i.e.,][]{Muzzin14, Belli19, Cheng25, Beverage25}. Fortunately, \verb|alf|$\alpha$ has additionally adopted the sMILES \citep{Knowles21, Knowles23} SSPs, which incorporate models as young as 0.03 Gyr, making them ideal for studying these galaxies. Furthermore, sMILES grids span $-1.5 < [\rm Z/H] < +0.26$ ratios, allowing for analyses that span a large range in total metallicities. From this, we can infer [Fe/H] using Equation 2 from \citet{Knowles23}. 

sMILES also spans $\rm -0.2 < [\alpha/Fe] < +0.6$ ratios, which is important, as the galaxies in this study are all recent PSB galaxies at high-$z$. They have likely not had time for SNe Ia to enrich Fe throughout the galaxy, and therefore should be $\alpha$-enhanced \citep[i.e.,][]{Kimmig25}. To do this, sMILES uses BaSTI solar-scaled isochrones \citep{BaSTI_I_solar} for the [$\alpha$/Fe]= $-0.20, \, 0.0, \, \& +0.20$ SSPs, and the BaSTI $\alpha$-enhanced isochrones \citep{BaSTI_II_alpha} for the [$\alpha$/Fe]=$+0.40 \, \rm and \, +0.60$ SSPs \citep{Knowles23}. sMILES varies the $\alpha$ elements O, Ne, Mg, Si, S, Ca and Ti in lock-step \citep{Knowles21, Knowles23}.

To understand the systemic differences between stellar population parameters recovered between the \citet{Conroy18} grids and sMILES grids, we fit a mock galaxy spectrum with both libraries. We discuss this testing in detail in Appendix \ref{sec:AA Testing}. We find that there is a persistent systemic difference of $\approx0.1$ dex between the two grids for both [Fe/H] and [$\alpha$/Fe]. This is to be expected, as different isochrones alone would naturally lead to different model spectra. However, quantifying this offset informs of the significance of this difference. Similar model-dependent offsets have been shown in \citet{Jafariyazani25}, with an average offset of $\approx0.1$ dex between [$\alpha$/Fe] inferred from the MILES \citep{Vazdekis15} and \citet{Conroy18} models, and an average offset of $\approx0.3$ dex between inferred [Fe/H] values. This is more extreme than the difference we find between the sMILES and \citet{Conroy18} grids, but highlights the need for future work investigating such systematic model uncertainties.

We use \verb|alf|$\alpha$ with the sMILES grids, assuming a \citet{Kroupa01} IMF. We utilize \verb|emcee| \citep{ForemanMackey13} for posterior sampling, and provide priors in Appendix \ref{sec:priors}.
We mask out the Na D$\lambda5890,5896$ absorption feature, as well as the H$\alpha$+\textsc{[N ii]} complex for our fitting. 
We additionally mask out the \textsc{[O iii]}$\lambda\lambda4959,5007$ doublet for XMM-VID3-2457. 
As discussed in \citet{Beverage25}, \verb|alf|$\alpha$ estimates the continuum by fitting a Chebyshev polynomial to the data-to-model ratio, applied independently in wavelength `chunks'. Unless otherwise specified, the default mode is to add one extra degree to the polynomial for each 100 $\textrm{\AA}$ in the chunk, rounded up (i.e., \verb|alf|$\alpha$  would use a 6th degree polynomial for a chunk of size 550 $\textrm{\AA}$). \citet{Beverage25} notes that results are robust to the choice of polynomial order. Nevertheless, to ensure consistency across radial bins and to avoid continuum subtraction systematics, we use identical wavelength chunks for all radial extractions within a given galaxy. Finally, we visually inspect the continuum fit after each run to verify the fit quality.

We note that rotational effects could impact our fitting results, particularly for XMM-VID3-2457, which is a fast rotator \citep{Forrest25}. At larger radii, annular bins span greater rotational velocities, introducing additional spectral broadening that may bias measurements of stellar population parameters that are sensitive to line widths. This broadening is accounted for by the velocity dispersion fit, which assumes a Gaussian line-of-sight velocity distribution. Deviations from a true Gaussian profile are expected to be small and are therefore folded into the fit uncertainties and noise model.

\section{Results} \label{sec:results}

\subsection{Integrated Spectral Fits}

We present the spectral fitting results for the extracted 1D spectra within 1 R$_{\textrm{e}}$ for each MQG.  Table \ref{tab:Bagpipes_Re_results} lists the \textsc{Bagpipes} results for both SFH models, while Table \ref{tab:AA_Re_results} summarizes the \texttt{alf}$\alpha$ results.  
XMM-VID1-2075 and XMM-VID3-1120 are both dust-poor, with $\rm Av = 0.04^{+0.04}_{-0.03}$ and $0.07^{+0.04}_{-0.04}$, respectively, while XMM-VID3-2457 has an $\rm A_V = 0.34^{+0.03}_{-0.04}$. These measurements are consistent with results from \citet{Chang26}, which studied the dust continuum of these three galaxies using ALMA observations. They found that XMM-VID1-2075 and XMM-VID3-1120 are extremely dust-poor, with $\rm M_{dust}/M_{\ast} \lesssim 10^{-4}$. XMM-VID3-2457 is relatively dust-rich in the context of MQGs, with $\rm M_{dust}/M_{\ast} \sim 10^{-3}$, approximately two dex below the main sequence \citep{Chang26}.

\begin{table*}[!htb]
\centering
\begin{tabular}{l|cccccccc}
\multicolumn{8}{c}{\textsc{Bagpipes}} \\ 
\multicolumn{8}{c}{\citet{Iyer19} SFH} \\
\hline
Galaxy &
$z_{\mathrm{spec}}$ &
$\log(M_{\ast}/M_{\odot})$  &
$\sigma_{\rm \ast}$ (km s$^{-1}$) &
log($Z/Z_\odot$)&
$\mathrm{A_V}$ &
${\rm SFR}_{100}~(\rm M_\odot\,{\rm yr}^{-1})$ &
$t\mathrm{_{MW}}$\footnote{mass-weighted age} (Gyr)\\
\hline

XMM\text{-}VID1\text{-}2075 &
$3.4465^{+0.0002}_{-0.0002}$ &
$11.42^{+0.02}_{-0.01}$ &
$334^{+14}_{-15}$ &
$0.33^{+0.11}_{-0.10}$ &
$0.04^{+0.04}_{-0.03}$ &
$0.00^{+0.03}_{-0.00}$ &
$0.59^{+0.04}_{-0.04}$\\

XMM\text{-}VID3\text{-}1120 &
$3.4863^{+0.0003}_{-0.0003}$ &
$11.37^{+0.01}_{-0.01}$ &
$336^{+16}_{-13}$ &
$0.40^{+0.04}_{-0.03}$ &
$0.07^{+0.04}_{-0.04}$ &
$1.77^{+0.43}_{-0.41}$ &
$0.65^{+0.03}_{-0.02}$\\

XMM\text{-}VID3\text{-}2457 &
$3.4868^{+0.0002}_{-0.0002}$ &
$11.24^{+0.01}_{-0.01}$ &
$341^{+10}_{-12}$ &
$0.19^{+0.07}_{-0.07}$ &
$0.34^{+0.03}_{-0.04}$ &
$1.21^{+0.40}_{-0.34}$ &
$0.46^{+0.03}_{-0.03}$\\
\hline

\multicolumn{8}{c}{\citet{Leja19} SFH} \\
\hline

XMM\text{-}VID1\text{-}2075 &
Fixed &
$11.42^{+0.01}_{-0.01}$ &
$328^{+14}_{-13}$ &
$0.19^{+0.07}_{-0.08}$ &
$0.04^{+0.05}_{-0.03}$ &
$0.05^{+0.74}_{-0.05}$ &
$0.55^{+0.04}_{-0.03}$\\

XMM\text{-}VID3\text{-}1120 &
Fixed &
$11.36^{+0.01}_{-0.01}$ &
$334^{+11}_{-12}$ &
$0.35^{+0.02}_{-0.03}$ &
$0.09^{+0.03}_{-0.04}$ &
$1.30^{+0.91}_{-0.45}$ &
$0.60^{+0.03}_{-0.02}$\\

XMM\text{-}VID3\text{-}2457 &
Fixed &
$11.22^{+0.01}_{-0.01}$ &
$339^{+12}_{-14}$ &
$0.19^{+0.05}_{-0.05}$ &
$0.43^{+0.03}_{-0.04}$ &
$5.52^{+4.41}_{-2.31}$ &
$0.34^{+0.02}_{-0.02}$\\
\hline

\end{tabular}
\caption{Posterior medians and 1$\sigma$ uncertainties from \textsc{Bagpipes}, assuming a flexible, non-parametric Gaussian Process SFH model \citep{Iyer19} (top three rows) and a smooth, continuity SFH model \citep{Leja19} (bottom three rows). Continuity model fits assume the median posterior redshift from \citet{Iyer19} SFH results. 
Fit results from both SFH models agree, with the exception of $\log(Z/Z_{\odot})$ and $t_{\rm MW}$. This confirms that our results are robust to choice of SFH prior.
Results for each galaxy are from our integrated ($\rm R < 1 R_{e}$) spectral extraction. Star formation rates are reported as the 100 Myr-averaged rate (SFR$_{100}$).
All three galaxies are very massive ($\mathrm{log(M_{_{\ast}} / M_{\odot})} > 11.2$), have large velocity dispersions ($\sigma_\ast \gtrsim 330$ km/s), are metal-rich ($\log(Z/Z_{\odot}) \gtrsim 0.2$), and have little-to-no dust ($\rm A_V \lesssim 0.4$). 
}
\label{tab:Bagpipes_Re_results}
\end{table*}

\begin{table*}[!htb]
\centering
\begin{tabular}{l|ccccc}
\multicolumn{6}{c}{\texttt{alf}$\alpha$} \\
\hline
Galaxy &
$t_{\rm SSP}$ (Gyr) &
[Fe/H] &
[Z/H] &
[$\alpha$/Fe] &
$\sigma_{\ast}$ (km s$^{-1}$) \\
\hline

XMM\text{-}VID1\text{-}2075 &
$0.63^{+0.06}_{-0.06}$ &
$-0.05^{+0.10}_{-0.12}$ &
$+0.04^{+0.11}_{-0.11}$ &
$+0.13^{+0.09}_{-0.08}$ &
$332^{+16}_{-15}$ \\

XMM\text{-}VID3\text{-}1120 &
$0.69^{+0.09}_{-0.04}$ &
$+0.06^{+0.07}_{-0.08}$ &
$+0.11^{+0.09}_{-0.13}$ &
$+0.08^{+0.08}_{-0.20}$ &
$353^{+14}_{-13}$ \\

XMM\text{-}VID3\text{-}2457 &
$0.53^{+0.03}_{-0.04}$ &
$+0.04^{+0.08}_{-0.09}$ &
$+0.03^{+0.09}_{-0.09}$ &
$-0.02^{+0.05}_{-0.05}$ &
$388^{+12}_{-12}$ \\
\hline
\end{tabular}
\caption{\texttt{alf}$\alpha$ fitting results, using $\alpha-$enhanced (sMILES) stellar population models. 
Ages are SSP-equivalent ($t_{\rm SSP}$). [Z/H] is the logarithmic abundance ratio of all metals relative to hydrogen, compared to the solar ratio. 
The MQGs are very young ($t_{\rm SSP}\lesssim 700$ Myr), and have metallicities that are consistent within $1\sigma$ of solar abundances.  XMM-VID1-2075 and XMM-VID3-1120 are both moderately $\alpha$-enhanced, while XMM-VID3-2457 is consistent with a solar [$\alpha$/Fe].
\texttt{alf}$\alpha$ also finds large velocity dispersions ($\gtrsim 330$ km/s) for all three MQGs, consistent with their extreme stellar masses.
}
\label{tab:AA_Re_results}
\end{table*}
 
Both \texttt{alf}$\alpha$ and \textsc{Bagpipes} show that the MQGs are young ($\rm {\it t}_{MW} < 0.7 \:Gyr$) and extremely massive ($\rm M_\ast\gtrsim 10^{11}\; \mathrm{M}_{\odot}$) for their epoch, similar to what has been shown for these MQGs in previous works using Keck/MOSFIRE spectroscopy \citep{Forrest20, Forrest22, Forrest24}. They also have large velocity dispersions ($\sigma_\ast \gtrsim 330$ km/s), consistent within 1$\sigma$ of previous measurements from \citet{Forrest22}.

We also determine quenching times ($t_q$) from the SFH. \textsc{Bagpipes} defines $t_q$ as when the following condition is satisfied in the galaxies history:

\begin{equation} \label{eq:nsfr}
    {\rm nSFR(t)} = \frac{t_{H} \times {\rm SFR}}{M_{\ast}} < 0.1
\end{equation}

where nSFR is the dimensionless normalized star-formation rate, $t_H$ is the age of the Universe, and $\rm M_{\ast}$ the mass formed by the galaxy by time $t_H$ \citep{Carnall18}. 
All three galaxies are recently quenched, with inferred quenching times of $\Delta t_q = t_{obs} - t_{q} = $ $0.24^{+0.03}_{-0.03}$ Gyr, $0.15^{+0.04}_{-0.06}$ Gyr, and $0.11^{+0.02}_{-0.07}$ Gyr before observation for XMM-VID3-1120, XMM-VID1-2075 and XMM-VID3-2457, respectively. 
The short separation between mass build-up and quenching ($t_{q} - t_{50} \lesssim 500$ Myr) indicates that they experienced rapid assembly just prior to quenching.
This is similar to what is shown in \citet{Carnall24} for comparable galaxies at a similar epochs, supporting a picture where high-$z$ massive quiescent systems undergo intense, short-lived starbursts followed by rapid quenching. 
We discuss this in further detail in Sec. \ref{sec:ages_masses}.

Given their extreme masses and compact nature ($\rm R_e \approx 2$ kpc), these MQGs have surface densities of $\Sigma_{\ast} = (1.8-2.6) \, \times 10^{10} \, \mathrm{M_{\odot} kpc^{-2}}$, comparable with the core densities of local ellipticals \citep[$\Sigma_{\ast} \approx10^{10}-10^{11} \, \mathrm{M_{\odot} kpc^{-2}}$;][]{Hopkins10, vanDokkum10, Mosleh17}. 
These high surface densities support a two-phase growth scenario, wherein these early, compact systems form the high-density cores of today's most massive ellipticals, with their subsequent structural growth occurring primarily through minor mergers \citep{Bezanson09, Oser10, Suess20}.

\subsubsection{Metallicities} \label{sec:metals}

The integrated spectra of XMM-VID1-2075 and XMM-VID3-1120 are moderately $\alpha$-enhanced ([$\alpha$/Fe] = $+0.13^{+0.09}_{-0.08}$ and $+0.08^{+0.08}_{-0.20}$, respectively). This is consistent with literature showing that post-starburst galaxies are $\alpha$-enhanced due to recent and rapid star-formation \citep[i.e.,][]{Kriek16, Beverage24, Beverage25, Kimmig25, Gountanis25, McConachie25, Hamadouche26}. However, these [$\alpha$/Fe] values are relatively low compared to other MQGs observed in literature, which generally exceed $\rm[\alpha/Fe]\gtrsim+0.3$ \citep[i.e.,][]{Kriek16, Beverage24, Beverage25, Gountanis25, McConachie25, Hamadouche26}. Furthermore, XMM-VID3-2457 is consistent within 1$\sigma$ of a solar abundance pattern ([$\alpha$/Fe]=$-0.02^{+0.05}_{-0.05}$). While this is somewhat surprising, \citet{Hamadouche26} observe a spread in  [$\alpha$/Fe] abundance ratios ($\langle [\rm \alpha/Fe] \rangle = +0.22^{+0.22}_{-0.17}$) for MQGs at this epoch. This may be due to early AGN outflows. As the ISM would not yet have been enriched in Fe-peak elements by SNe Ia, these outflows would preferentially eject $\alpha$-elements, which would have already been present. This would naturally lower the observed [$\alpha$/Fe].

Additionally, \citet{Hamadouche26} notes that while SNe Ia have long enrichment timescales, their delay-time distribution (DTD) after a star-formation event follows a power law that is $\rm DTD (\tau) \propto t^{-1}$ \citep{DeLucia14}. This allows for the onset of chemical enrichment from SNe Ia to occur as early as $\sim40$ Myr after such a star formation event, the minimum time for white dwarfs to form. Fe-peak elements would then be enriched earlier than expected. The [$\alpha$/Fe] abundance ratio in the galaxy would then be correspondingly reduced once the Fe-enriched gas is incorporated into new stars. For more discussion on this, see \citet{Hamadouche26}. 
Alternatively, prior SNe Ia enrichment before the recent burst of star-formation could have enriched the stellar population with Fe-peak elements. This is consistent with the supersolar metallicities reported from \textsc{Bagpipes}. 

Disconcertingly, the metallicities inferred from \textsc{Bagpipes} and \texttt{alf}$\alpha$ are offset by $\approx0.2-0.4$ dex (Tables \ref{tab:Bagpipes_Re_results} \& \ref{tab:AA_Re_results}).
Recent literature has found similar offsets between codes using solar-scaled libraries and those that allow for variable elemental abundances. \citet{Beverage25} compared the metallicities from \texttt{Prospector} \citep{Leja19, Johnson21}, which uses solar-scaled stellar libraries, to metallicities from \texttt{alf}$\alpha$, using the \citet{Conroy18} grid to vary individual elemental abundances.  They found a large scatter, with offsets up to $\approx 0.7$ dex. Similarly, \citet{McConachie25} found a $\approx 0.15$ dex offset between \verb|Prospector| and \verb|alf|$\alpha$ metallicities. \citet{Leung26} confirmed this is also the case between \textsc{Bagpipes} and \texttt{alf}$\alpha$ (using the sMILES grids), finding offsets as large as $\approx 0.7$ dex. While the disagreement between the metallicities found in this work are less severe than what is presented in other literature, this emerging trend is concerning, and future work is needed to reconcile it.

A secondary effect that can contribute to this offset is the difference in zero-point solar abundances assumed between \textsc{Bagpipes} and \texttt{alf}$\alpha$, based on the choice of stellar library.
\textsc{Bagpipes} utilizes the 2016 version of the \citet{BC03} stellar population models, which adopts the MILES \citep{MILES2011, Vazdekis15} library of empirical spectra. MILES assumes a solar metallicity of Z$_{\odot} = 0.0198$ from \citet{GN93}. sMILES uses the BaSTI isochrones, which in turn also assumes the \citet{GN93} value for Z$_{\odot}$. However, sMILES adopts the \citet{Asplund05} solar abundances for their theoretical stellar spectra, which has Z$_{\odot} = 0.0122$. Equation 2 in \citet{Knowles23} accounts for this by converting their SSP values to the \citet{GN93} abundances so that they are using the right isochrone. Importantly, \citet{Knowles23} notes that this conversion is not fully consistent, resulting in different $\alpha$-element abundances. However, they note that this is a secondary effect to the choice in stellar spectra to use \citep{Vazdekis15}.

\subsubsection{Ages \& Mass Build-up} \label{sec:ages_masses}

\begin{figure*}[!bht]
\includegraphics[width=\textwidth]{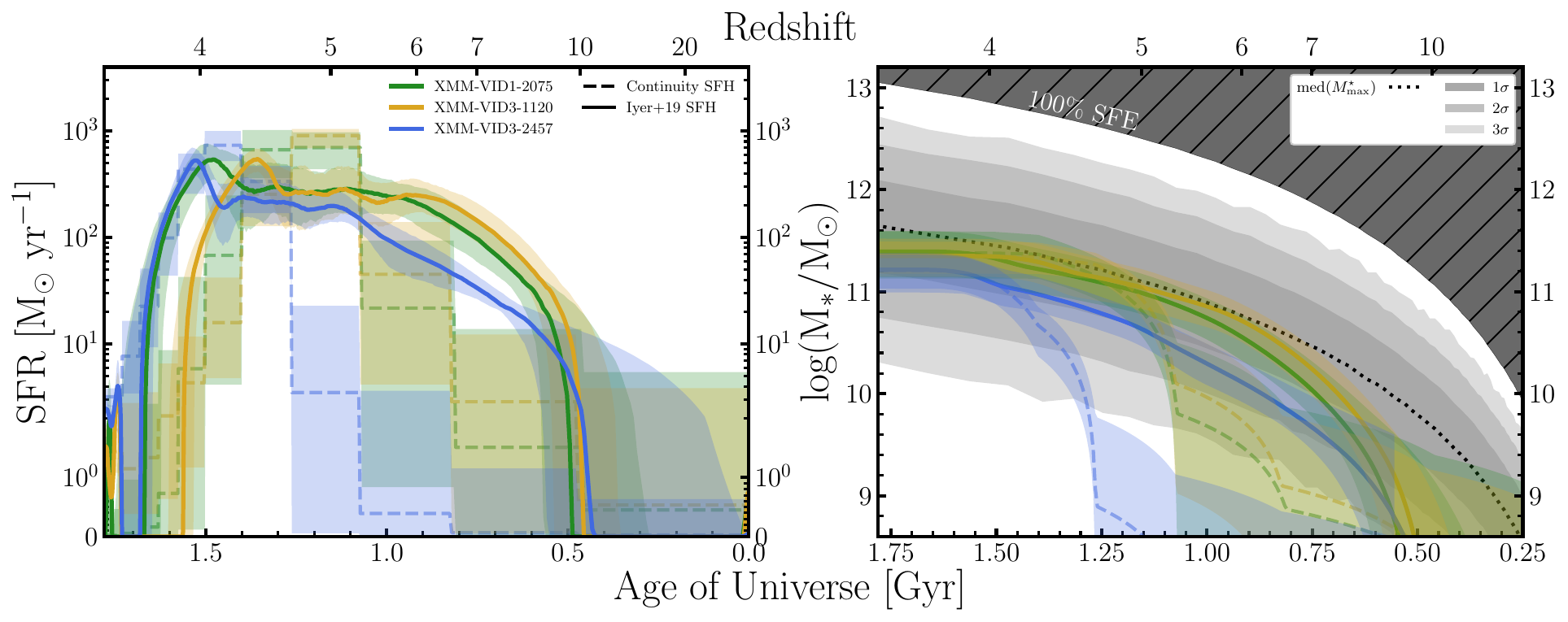}
\centering
\cprotect\caption{\textit{Left}: \textsc{Bagpipes} star-formation histories. Solid lines show the non-parametric Gaussian Process SFH \citep{Iyer19}, and dashed lines show SFHs adopting the continuity model \citep{Leja19}. \textit{Right}: Mass assembly histories based on the inferred SFHs. We compare these with the probability density function of the most massive galaxy expected in the XMM-VIDEO survey volume (assuming an area of 4.65 deg$^{2}$), computed using \texttt{evstats} \citep{Lovell23}. The hatched region corresponds to where \texttt{evstats} requires $100\%$ SFE at a $>3\sigma$ confidence level.
All three galaxies formed early, reaching masses of $\gtrsim 10^{10} \mathrm{M}_{\odot}$ in the first $\sim 800$ Myr of cosmic time ($z\sim 6.8$). XMM-VID1-2075 and XMM-VID3-1120 both are approaching $\rm M_{\ast} \sim 10^{11} \mathrm{M}_{\odot}$ by $z\sim 6$.
}
\label{fig:Re_SFH_MAH}
\end{figure*}

We present the SFH and Mass Assembly History for each galaxy in Figure \ref{fig:Re_SFH_MAH}. All three MQGs experienced a recent, intense starburst, with star-formation rates exceeding $500~{\rm M}_{\odot}~\mathrm{yr}^{-1}$ within the $500$ Myr prior to observation, corresponding to $4 \lesssim z \lesssim 5$. 
All three MQGs also assembled very early. XMM-VID3-2457 assembled $\gtrsim10\%$ of its total formed stellar mass ($\rm M_\ast \approx 10^{10} M_{\odot}$) by $z\sim6$, while XMM-VID1-2075 and XMM-VID3-1120 both assembled $\gtrsim20\%$ of their total formed stellar mass ($\rm M_\ast \approx 10^{10.8} M_{\odot}$) during this same time.
The latter two nearly reach the ultra-massive galaxy regime ($\rm M_{\ast} \gtrsim 10^{11} \mathrm{M}_{\odot}$) within the first $\sim1$ Gyr. 

While the masses reported by \textsc{Bagpipes} are extreme, they are consistent with the most massive galaxy expected in a survey the size of XMM-VIDEO ($\sim 4.65 \: \mathrm{deg}^{2}$), as quantified by \texttt{evstats} \citep{Lovell23}. 
\texttt{evstats} computes the probability density function of the most massive galaxy expected in the XMM-VIDEO survey volume ($4.65\rm \, deg^{2}$), assuming a baryon fraction $f_b = 0.16$ and a $\Lambda$CDM cosmology \citep{Lovell23}.

The GP-SFHs agree with the continuity fits at recent times. At early times ($z\gtrsim5$) however, the continuity model does not capture the predicted early mass growth from the GP-SFHs. We note that the amount of early mass growth is highly sensitive to the choice of scale factor in the continuity prior. Changing this to $\sigma=0.3$ results in excess mass in the early bins relative to what is physically allowed in $\Lambda$CDM. Alternatively, $\sigma=1.0$ results in most of the mass forming in 1-3 bins, consistent with the peaks observed in the GP-SFHs. We note that the fitting results from the continuity models are consistent within 1-2$\sigma$ with the GP-SFH results.
Therefore, our main results are robust to the choice of SFH prior.

The choice of IMF can also impact inferred stellar masses. \citet{Cheng26} showed that early, extremely massive galaxies could prefer a bottom-heavy IMF such as the \citet{Salpeter55} IMF. In contrast, \textsc{Bagpipes} utilizes a \citet{KroupaBoily02} IMF, which would infer masses about 3-4 times less massive than the \citet{Salpeter55} IMF \citep{Cheng26}. Our mass measurements are therefore conservative. Moreover, they are realistic for our sample, as \citet{Forrest22} found that the dynamical masses for these MQGs prefers a `bottom-light' IMF like \citet{Chabrier03}.

\subsection{Resolved Spectral Fits}
\subsubsection{\texorpdfstring{Age, Metallicity, and $[\alpha/\rm Fe]$ Gradients}{Age, Metallicity, and [a/Fe] Gradients}} \label{sec:gradients}

\begin{figure*}[!bthp]
\includegraphics[width=\textwidth]{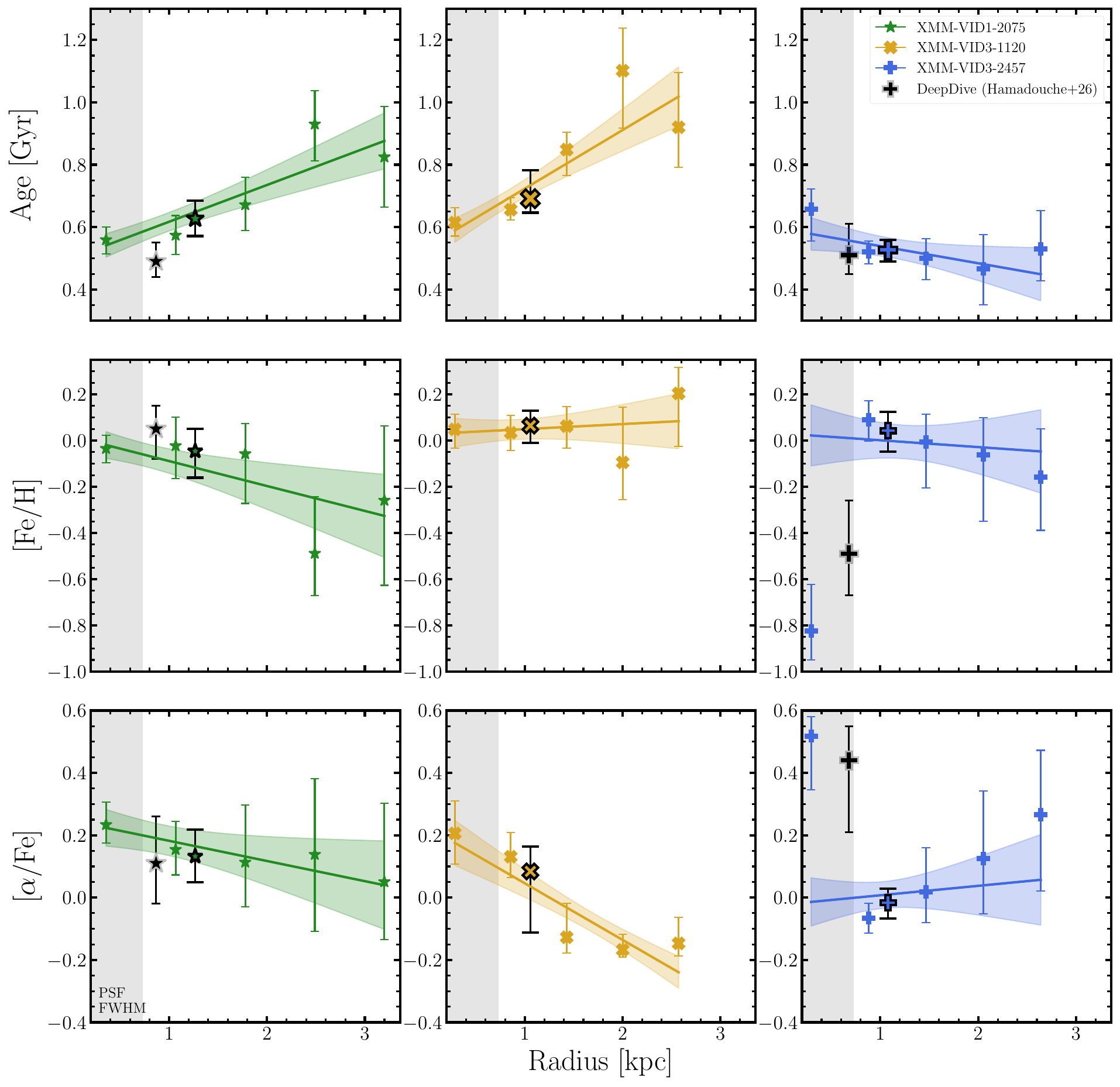}
\centering
\cprotect\caption{Age (\textit{Top}), [Fe/H] (\textit{Middle}), and [$\alpha$/Fe] (\textit{Bottom}) gradients for all three galaxies in this study. We outline measurements within $\rm R < 1 R_{e}$ and plot them at $\frac{1}{2}\rm R_e$, slightly shifted for visual purposes. The remaining points are measured in elliptical annuli centered around the galaxy, in bins with a semi-major axis width of 1 pixel. We illustrate gradients using a least-squares fit to the radial measurements. Additionally, we plot results from \citet{Hamadouche26}, which observed two of our galaxies using NIRSpec MSA slit spectroscopy, as black points with a gray outline. The FWHM of the PSF at the red end of the spectrum ($\sim1$ pixel, or $\rm FWHM_{PSF}\approx730$pc) is shown in gray.}
\label{fig:Radial_grad}
\end{figure*}
\begin{figure}[!hbtp]
\includegraphics[width=\columnwidth]{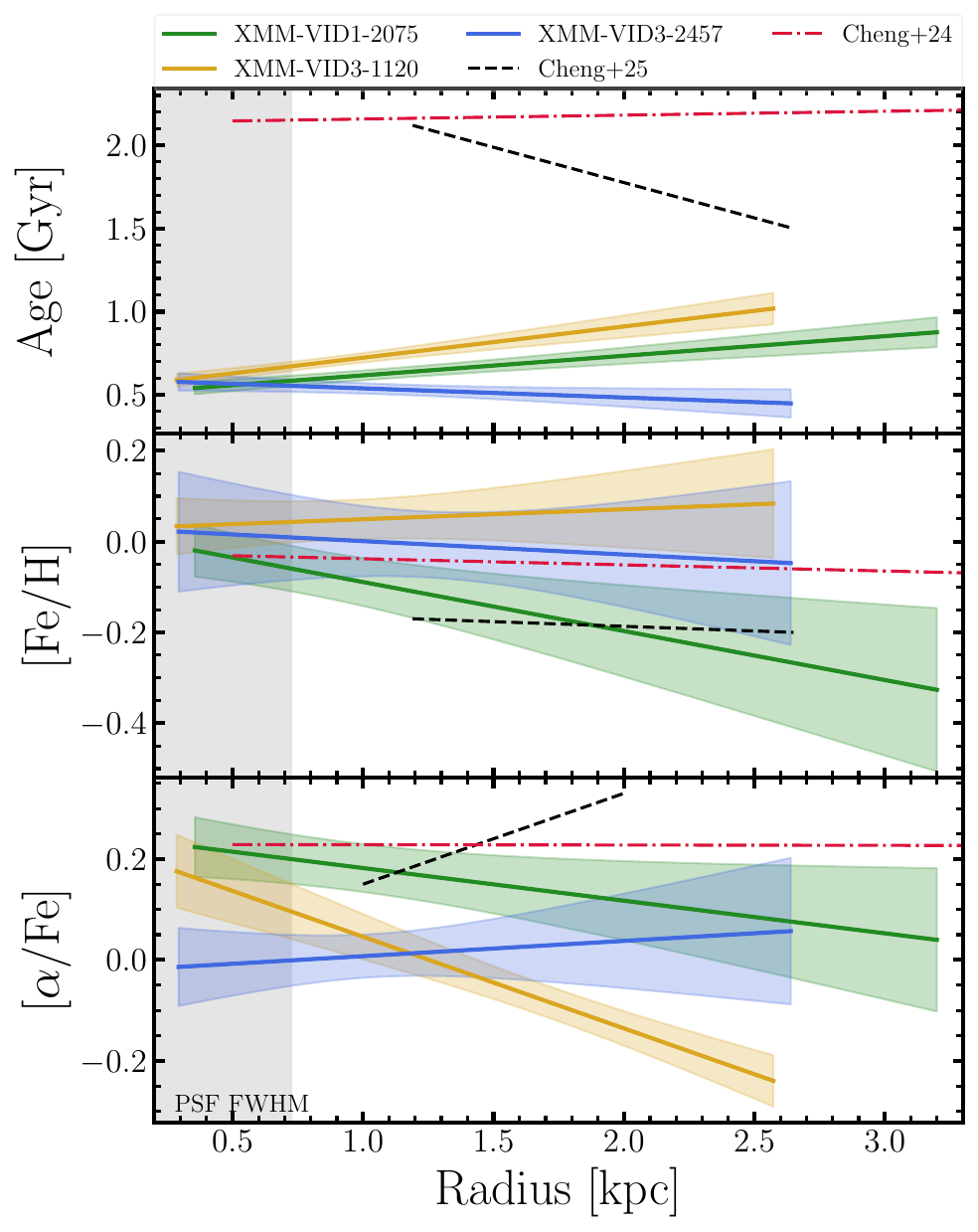}
\centering
\cprotect\caption{Same as Fig. \ref{fig:Radial_grad}, but with all galaxies plotted together for visual comparison. Data points are not shown for clarity, but we instead show the trend lines and associated uncertainties. We additionally show the median gradients from \citet{Cheng25} for massive quiescent galaxies at $1.2 < z < 2.2$ in JWST-SUSPENSE as a black dashed line for comparison. We also show the median gradients from the `young quiescent' LEGA-C galaxies at $0.6 < z < 1$ in \citet{Cheng24} as a red dash-dotted line.}
\label{fig:Radial_grad_all}
\end{figure}

We present radial gradients of age, [Fe/H], and [$\alpha$/Fe] for each galaxy in Figure \ref{fig:Radial_grad}, and all together in Figure \ref{fig:Radial_grad_all}. 
We compare our results to those from JWST/NIRSpec slit spectra of the same sources from the \textit{DeepDive} survey \citep{Hamadouche26}, which observed two of our three targets.

The gray shaded region represents the expected full width half max (FWHM) of the point spread function (PSF) at the red end of the spectrum, measured using \texttt{StPSF}. At the blue end, the PSF FWHM corresponds to $\approx \frac{1}{2}$ pixel, or $\approx 370$ pc at the redshift of the MQGs. At the red end of the spectrum, it corresponds to $\approx$ 1 pixel, or $\approx 730$ pc. We note that this change in PSF would flatten our derived stellar population gradients. Therefore, while a correction would result in steeper gradients, the directionality of the gradients would not change. Hence, our main results are robust against PSF effects.

We note that our integrated IFU measurements are consistent with those derived from MSA slit spectroscopy, although the IFU yields smaller uncertainties. 
However, in XMM-VID3-2457, the MSA-based results for [Fe/H] and [$\alpha$/Fe] better resemble IFU-based results from the core, rather than from the integrated measurements.
The most likely cause of this is that the central portion of the galaxy dominates the light of the slit spectra. 
The slit width of the MSA is $0.2^{\prime\prime}$, while XMM-VID3-2457 has an effective radius $\rm R_e=0.28.^{\prime\prime}$
Therefore, the integrated IFU results present a more complete look at the entire stellar population of these galaxies.

Our full spectrum fitting with {\tt alf$\alpha$} reveals that the cores ($\rm R \lesssim 0.5$ kpc) of each galaxy are young ($t_{\rm SSP} \lesssim 700$ Myr) and $\alpha$-enhanced ([$\alpha$/Fe] $\gtrsim +0.2$), with XMM-VID1-2075 and XMM-VID3-1120 exhibiting solar metallicities ($\rm [Fe/H] = -0.03^{+0.06}_{-0.06}$ and $\rm [Fe/H] = +0.05^{+0.07}_{-0.08}$, respectively). This implies that they formed via a similar process. The young ages and elevated [$\alpha$/Fe] indicate rapid, recent star-formation, suggesting a past starburst \citep[e.g.,][]{Kimmig25}. Post-starburst regions show similar SFHs and analogous chemical enrichment properties, regardless of spatial location within a galaxy \citep{Leung25}, suggesting that while the direct causes of the starburst might vary, they dominate local processes in the galaxy and result in comparable stellar populations. An extreme central starburst is therefore a likely explanation for these observations.

Beyond the cores, both XMM-VID1-2075 and XMM-VID3-1120 have positive age gradients and negative [$\alpha$/Fe] gradients. XMM-VID1-2075 also has a negative [Fe/H] gradient, while XMM-VID3-1120 has a roughly flat [Fe/H] gradient. 
XMM-VID3-2457, on the other hand, displays a flat age gradient. Outside of the central bin, it has negative [Fe/H] and positive [$\alpha$/Fe] gradients. 
It would appear that XMM-VID1-2075 and XMM-VID3-1120 have likely experienced an intense central starburst followed by strong feedback, causing rapid quenching. 
XMM-VID3-2457 seems to have quenched uniformly, plausibly by an AGN given the presence of \textsc{\textsc{[O iii]}}$\lambda\lambda4959,5007$ and \textsc{\textsc{[N ii]}}$\lambda\lambda6548,6584$ emission.
Further discussion on the implications of these gradients can be found in Section \ref{sec:UMG_formation}.

We again note that rotational effects are a concern at larger radii, particularly for XMM-VID3-2457, which is a fast rotator \citep{Forrest25}. To account for potential line broadening, we include velocity dispersion as a free parameter in our {\tt alf$\alpha$} fits. Although we assume a Gaussian shape, any offsets resulting from this functional form and the intrinsic line shape are small and fully accounted for in the noise modeling. Future work will more thoroughly test this using velocity dispersion profiles. However, we note that the outermost bins are consistent with their respective trends, which is encouraging.

These gradients highlight the importance of IFU spectroscopy in studying the evolution of massive quiescent galaxies. The UVJ color evolution of these MQGs are roughly flat, with mean slopes of $\rm \langle \frac{d(UV)_{rest}}{d\,kpc} \rangle = -0.02$ and $\rm \langle \frac{d(VJ)_{rest}}{d\,kpc} \rangle = 0.04$. As noted in \citet{Cheng24}, broad-band colors and low-resolution spectra would be unable to break the age-dust-metallicity degeneracy for such galaxies. However, with the high resolution spectra available from JWST/NIRSpec IFU, we can do so using absorption line diagnostics, allowing us to measure physically meaningful gradients.

\subsubsection{Resolved Star Formation Histories}\label{sec:rad_SFH} 
\begin{figure*}[!htb]
\includegraphics[width=\textwidth]{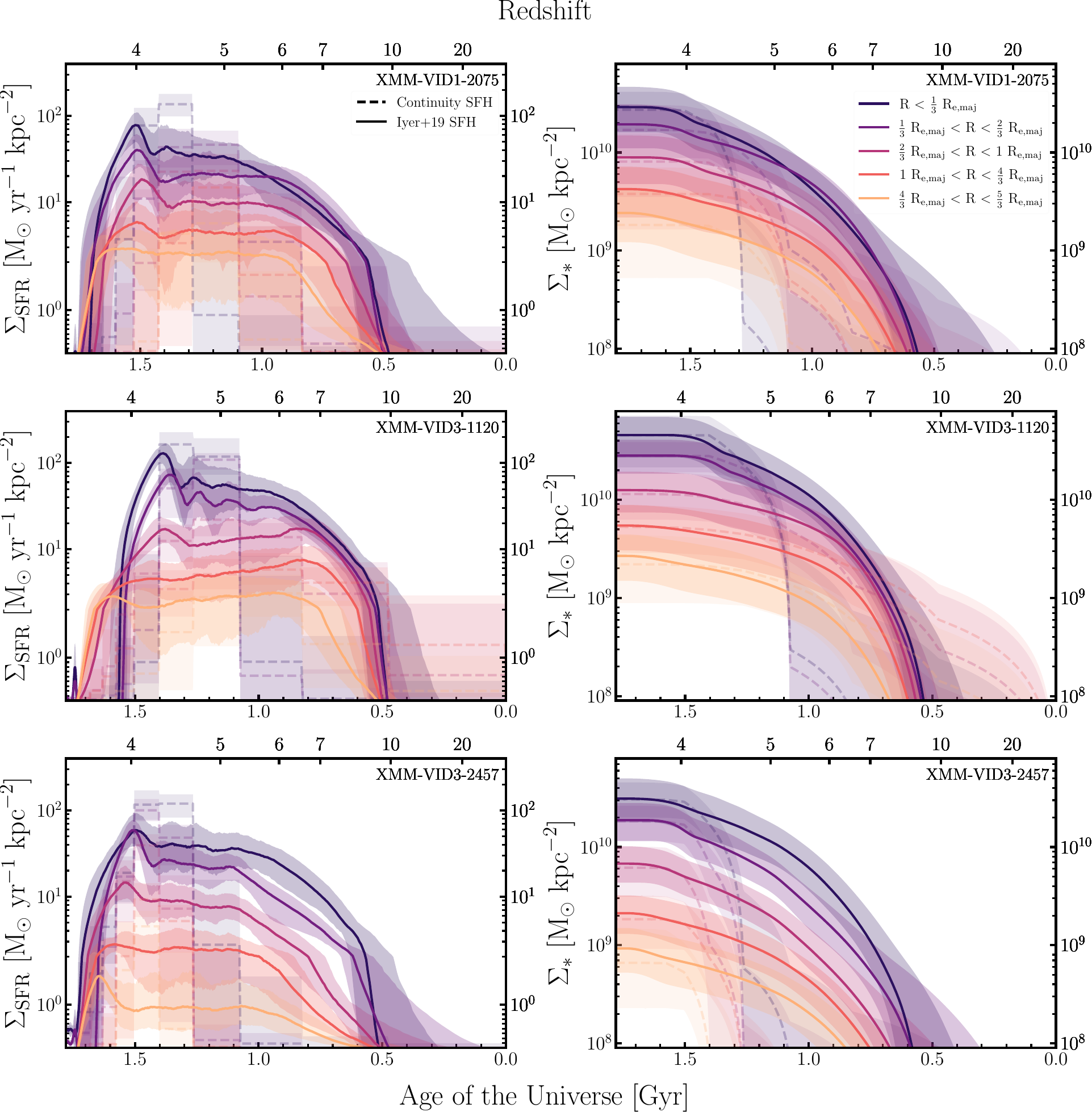}
\centering
\cprotect\caption{\textit{Left}: Radial SFHs of each massive quiescent galaxy, modelled by \textsc{Bagpipes} assuming a Gaussian Process SFH \citep{Iyer19} (solid lines) and continuity model \citep{Leja19} (dashed lines). \textit{Right}: Mass assembly history as a function of cosmic time. Both are shown in units of surface density. Shaded regions correspond to $1\sigma$ confidence intervals. 
}
\label{fig:UMG_SFH_MAH_radial}
\end{figure*}

We present the SFHs for each elliptical annular bin in Figure \ref{fig:UMG_SFH_MAH_radial}.
The left panels show the SFHs, modeled using \textsc{Bagpipes}, as described in Section \ref{Sec: Bagpipes}. 
The right panels show the mass assembly history, calculated from the SFH. Both are in units of surface density. Like the integrated SFHs, the continuity and GP-SFHs are in good agreement near the time of observation. Above $z\gtrsim5$, however, the continuity model does not predict the early mass growth the GP-SFH models do, similar to the disagreement seen in the integrated SFHs (Section \ref{sec:ages_masses}). The radial SFHs for all three MQGs are consistent with an extreme central starburst ($\rm \Sigma_{SFR} \gtrsim 50 \, M_{\odot} \, yr^{-1} \, kpc^{-2}$) occurring between $4 < z < 5$, and substantial early mass growth above $z\sim6$, similar to our integrated results. 
They also illustrate a more complicated picture of past quenching of the MQGs than previous observations \citep{Forrest20}.

All radial bins of XMM-VID3-2457 quench at the same time within uncertainties ($t_{q, \, core} - t_{q, \, outskirts} = 60^{+44}_{-44}$ Myr), consistent with the pattern seen in the age gradients. XMM-VID3-1120 quenches in an inside-out pattern, where the central bins quench earlier than the outskirts ($t_{q, \, core} - t_{q, \, outskirts} = 128^{+55}_{-41}$ Myr). Similar to XMM-VID3-2457, XMM-VID1-2075 quenches its annular bins at the same time within uncertainties ($t_{\rm q, core} - t_{\rm q, outskirts} = 25^{+66}_{-83}$ Myr).
This is puzzling, as this is counter to what is expected from the positive age gradient seen in XMM-VID1-2075 from \texttt{alf}$\alpha$ fitting.

This difference is likely due to how the different stellar codes treat $\alpha$-elements. 
For example, the inner-most bin of XMM-VID3-2457 is the most $\alpha$-enhanced spectrum in our sample with $[\alpha\rm /Fe]=+0.52^{+0.06}_{-0.17}$, which would indicate a formation time $t_{\rm form} = 12.4^{+117.1}_{-2.9}$ Myr using the relationship $[\alpha/{\rm Fe}] = \frac{1}{5} - \frac{1}{6}\log_{10}(\Delta t_{\rm form} / {\rm Gyr})$ \citep{Thomas05}. In comparison, \textsc{Bagpipes} SFH modeling shows that the formation timescale, $\tau_{\rm SF} =t_{80}-t_{20} = 423\pm68$ Myr. This is expected, as \textsc{Bagpipes} uses solar-scaled stellar libraries, which do not consider how $\alpha$-enhancement can shorten formation timescales. Future work using $\alpha$-MC \citep{Park25} or incorporating sMILES into \textsc{Bagpipes} as done in \citet{Leung26} will be important to better constrain these timescales. Furthermore, as noted in \citet{Leung26b}, making a distinction between `inside-out', `outside-in', and `uniform' quenching is perhaps overly simplistic when compared to complex, spatially resolved SFHs. We discuss this further in Section \ref{sec:QuenchingSequence}. 

\section{Discussion} \label{sec:discuss}

\subsection{Formation Processes for MQGs at Early Times} \label{sec:UMG_formation}

\subsubsection{Central Starburst}

In this paper, we have presented a spatially-resolved study of massive quiescent galaxies at $z > 3$. Our three targets are dominated by post-starbursts with remarkably similar stellar population properties. This homogeneity implies a potential common evolutionary pathway that formed the core of all three MQGs \citep{Leung25}. The young ($t_{\rm SSP}\lesssim 700$ Myr) ages and high [$\alpha$/Fe] ($\gtrsim +0.2$) in these regions are consistent with a short but intense central starburst. Such a starburst would allow for significant SNe II enrichment but little SNe Ia enrichment, elevating $\rm [\alpha/Fe]$. This is reinforced by the radial SFHs (Fig. \ref{fig:UMG_SFH_MAH_radial}), which show that the cores all had extreme ($\rm \Sigma_{SFR} > 50 \, M_{\odot} \, yr^{-1} \, kpc^{-2}$) star-formation rate densities at $4 < z < 5$.

The extreme star-formation rates inferred by the SFHs suggest an evolutionary link to sub-millimeter galaxies (SMGs) at $4 \lesssim z \lesssim 6$ as the progenitors of MQGs at $3<z<4$ 
\citep[e.g.,][]{Toft14, Marchesini14, Wilkinson17, GomezGuijarro18, Forrest20a, Valentino20, Carnall23MQG, Glazebrook24, Skarbinski25, PAA25b, Zavala26}. The inferred extreme star-formation rates and stellar masses of the MQGs between $4 \lesssim z \lesssim 6$ agree within $\sim1\sigma$ of those reported for a sample of SMGs in COSMOS at this epoch \citep[$\rm SFR\gtrsim500\, M_{\odot \, }yr^{-1}, \, M_{\ast}\approx10^{11}M_{\odot} $;][]{Smolcic15}, hinting further that SMGs may be their progenitors.

Recent theoretical work corroborates this picture. \citet{PAA25b} uses a recalibrated version of the \texttt{L-Galaxies} Semi-Analytical Model \citep{PAA2025a} that reproduces the observed number densities of both massive quiescent galaxies and SMGs. In their model, MQGs in the ultra-massive regime ($\rm M_{\ast} > 10^{11}M_{\odot}$) were all sub-mm bright (S$_{870} \gtrsim 1$mJy) at some point in their past.  
During the sub-mm bright phase, the MQGs experienced major mergers and black hole growth. As the gas reservoir empties from the starburst induced by the merger, AGN feedback stops further gas cooling \citep{PAA25b}. They find that $z=3$ and $z=4$ MQGs quench 0.6 Gyr and 0.4 Gyr after the peak sub-mm phase occurs. This is shorter than typical dust destruction timescales \citep[$\sim$100 Myr; e.g.,][]{Jones11}, which implies that their descendants would be dust-poor when observed. This is the case for our sample, as they are dust-poor (A$_{\rm V} \lesssim 0.4$). ALMA Band 7 observations confirm this, as two are undetected, and XMM-VID3-2457 has only a faint 870$\mu$m dust continuum detection \citep{Chang26}. 


\subsubsection{Difference in Gradients} \label{sec:grad_diff}

We can see from Figure \ref{fig:Radial_grad} that the MQGs exhibit different radial trends. To compare these radial trends, we construct nested models to quantify their independence. 

We first build model M0, where each measurement parameter (age, [Fe/H], [$\alpha$/Fe]) is fit with a single slope and intercept as a function of radius, fit using measurements from all three galaxies. 
We then build model M1, where each parameter for each MQG is assigned a slope that is shared among all three galaxies, but a unique intercept for each. 
We finally build model M2, where the data for each galaxy are fit with an independent slope and intercept. 

M2 would be preferred in a scenario where the MQGs have statistically different radial trends, both in intercept and slope for a given parameter. This would be expected in a scenario where the MQGs experienced significantly divergent evolutionary pathways. M0, alternatively, would be preferred if the MQGs exhibited similar gradients across all samples, implying they experienced near identical evolutionary effects throughout their volumes. 
M1 is preferred when the MQGs share a statistically similar slope but differ in normalization, implying the MQGs would have experienced similar processes with varying strengths or timescales.

We compare each model for each parameter using a $\Delta$BIC test \citep{Schwarz78}, where the Bayesian Information Criterion (BIC) is defined as:

\begin{equation}
    \mathrm{BIC} = k \, \mathrm{ln}(n) -2L
\end{equation}

where $k$ is the number of parameters in a given model, $n$ is the sample size, and $L$ is the log-likelihood evaluated for the model. 
This provides a statistical test of how well the model fits the data, while penalizing overfitting.
When comparing two models, the model with the lower BIC is preferred. 
$0 < |\Delta\mathrm{BIC}| \le 2$ is considered weak evidence in favor of the model with a lower BIC, $2 < |\Delta\mathrm{BIC}| \le 6$ is considered moderate evidence, $6 < |\Delta\mathrm{BIC}| \le 10$ is considered strong evidence, and $10 < |\Delta\mathrm{BIC}|$ is considered very strong evidence \citep{Kass95}.

We use \texttt{statsmodels} \citep{statsmodels} to evaluate the log-likelihood for each model. For each parameter, M2 is found to have the lowest BIC, and M1 usually the second lowest. The only exception is for [Fe/H], where M0 has a $\rm BIC = -8.8$ and M1 has $\rm BIC = -8.4$. M2 is preferred strongly for age ($\Delta$BIC = 9.2) and moderately for [Fe/H] ($\Delta$BIC = 2.5) and [$\alpha$/Fe] ($\Delta$BIC = 5.8) when compared to M1. When we exclude XMM-VID3-2457, we find that M1 is weakly preferred for both age ($\Delta$BIC = 1.8) and [Fe/H] ($\Delta$BIC = 0.1), and M2 is strongly preferred for [$\alpha$/Fe] ($\Delta$BIC = 10.7). We therefore conclude that there is moderate to strong statistical evidence that XMM-VID3-2457 has statistically distinct radial trends than XMM-VID1-2075 and XMM-VID3-1120, both of which appear to have similar age and [Fe/H] trends. This would suggest that XMM‑VID3‑2457 has a distinct quenching pathway, while XMM-VID1-2075 and XMM-VID3-1120 seem to have undergone similar processes.

\subsubsection{Inside-out vs. Uniform Quenching} \label{sec:QuenchingSequence}

We can quantitatively say from Section \ref{sec:grad_diff} that XMM-VID1-2075 and XMM-VID3-1120 exhibit statistically similar radial trends. XMM-VID1-2075 has a positive age gradient, a negative [Fe/H] gradient, and a mildly negative [$\alpha$/Fe] gradient. XMM-VID3-1120 exhibits a positive age gradient and a flat [Fe/H] gradient. It also has a very negative [$\alpha$/Fe] gradient, such that the outer radial bins push against the lower limit of the sMILES grids ($[\rm \alpha/Fe]=-0.2$). The flat [Fe/H] gradient suggests relatively uniform Fe enrichment across the galaxy, consistent with widespread SNe Ia enrichment prior to the central starburst which enriched the core with $\alpha$-elements.

Both galaxies have cores that are young and $\alpha$-enhanced. This is consistent with a scenario where, after a major merger, gas is compacted to the center of the galaxy, igniting a central starburst. Strong AGN feedback then suppresses the depleted gas reservoir from forming stars in the central regions \citep{Park23, McClymont25, Kimmig25, PAA25b, Ni25}. This feedback propagates outwards, quenching the galaxy on short timescales \citep[$\sim100-200$ Myr;][]{DEugenio20, Park23, Kimmig25, McClymont25, Ni25}. This is unlike inside-out quenching, which occurs on much longer timescales \citep[$\gtrsim1$ Gyr; e.g.,][]{Walters22, Lawlor-Forsyth26}. Rapid, merger driven quenching results in positive age gradients, as the centers have younger light-weighted ages from the starburst \citep{Wu20, Pathak21, Ni25}, as seen in both XMM-VID1-2075 and XMM-VID3-1120. 

XMM-VID1-2075 shows evidence of such a major merger capable of triggering an extreme starburst, with extended low-surface brightness morphological asymmetries and a low ($\rm \lambda_{R_{e}} \approx 0.1$) spin parameter \citep{Forrest25}. It falls in the slow-rotator region of the spin parameter-ellipticity plane, very likely due to past major mergers \citep{Forrest25}. Although XMM-VID3-1120 is not a slow rotator, it has a low spin parameter ($\rm \lambda_{R_{e}} \approx 0.3$) compared to XMM-VID3-2457, which is a fast rotator. 
Therefore, based on its extreme past SFH (Fig. \ref{fig:Re_SFH_MAH}), it is likely that XMM-VID3-1120 also experienced a recent merger event, though further analysis of its kinematics is required to confirm this. An alternative explanation is that the starburst was the result of a significant influx of cold molecular gas, however this would require a very large gas supply for such an in-situ formation. Ex-situ growth from merger events would be more plausible \citep{Cochrane25}.

Unlike the others, XMM-VID3-2457 exhibits statistically different gradients (Sec. \ref{sec:grad_diff}). It has a flat age gradient, and is very Fe-deficient in its center relative to the outskirts. Beyond the core, it has a negative [Fe/H] gradient. Like XMM-VID3-1120, it is significantly alpha-enhanced in the core ([$\alpha$/Fe] = $+0.52^{+0.06}_{-0.17}$). This is likely caused by a massive, centrally contained starburst, enriching the core with $\alpha$-elements via SNe II. This could also be due to a wet merger, or accretion of cold, metal-poor gas. Both scenarios naturally explain the central Fe‑deficiency by diluting the pre‑existing gas. Star formation seems to have turned off uniformly, as evidenced by the flat age gradient and radial SFH, which shows that its star-formation ceased globally in a short amount of time. 
This is best explained by an internal mechanism, likely AGN feedback. The spectrum of XMM-VID3-2457 has evidence for an AGN, with strong \textsc{[O iii]}$\lambda\lambda$4959,5007 emission, as well as emission in the H$\alpha$+\textsc{[N ii]} complex, potentially indicating an AGN \citep{Hamadouche26}. While this provides the best explanation for the observed radial gradients, we have not necessarily confirmed the presence of one; \citet{Chang26} finds a relatively small AGN contribution ($f_{\rm AGN} = 0.16 \pm0.11$) for this MQG. This does not necessarily eliminate the possibility of an AGN quenching this MQG in the past, as AGN in post-starburst galaxies have short duty cycles \citep[$\sim 5.3\%$;][]{French23}.

\begin{figure*}[!htbp]
\includegraphics[width=\textwidth]{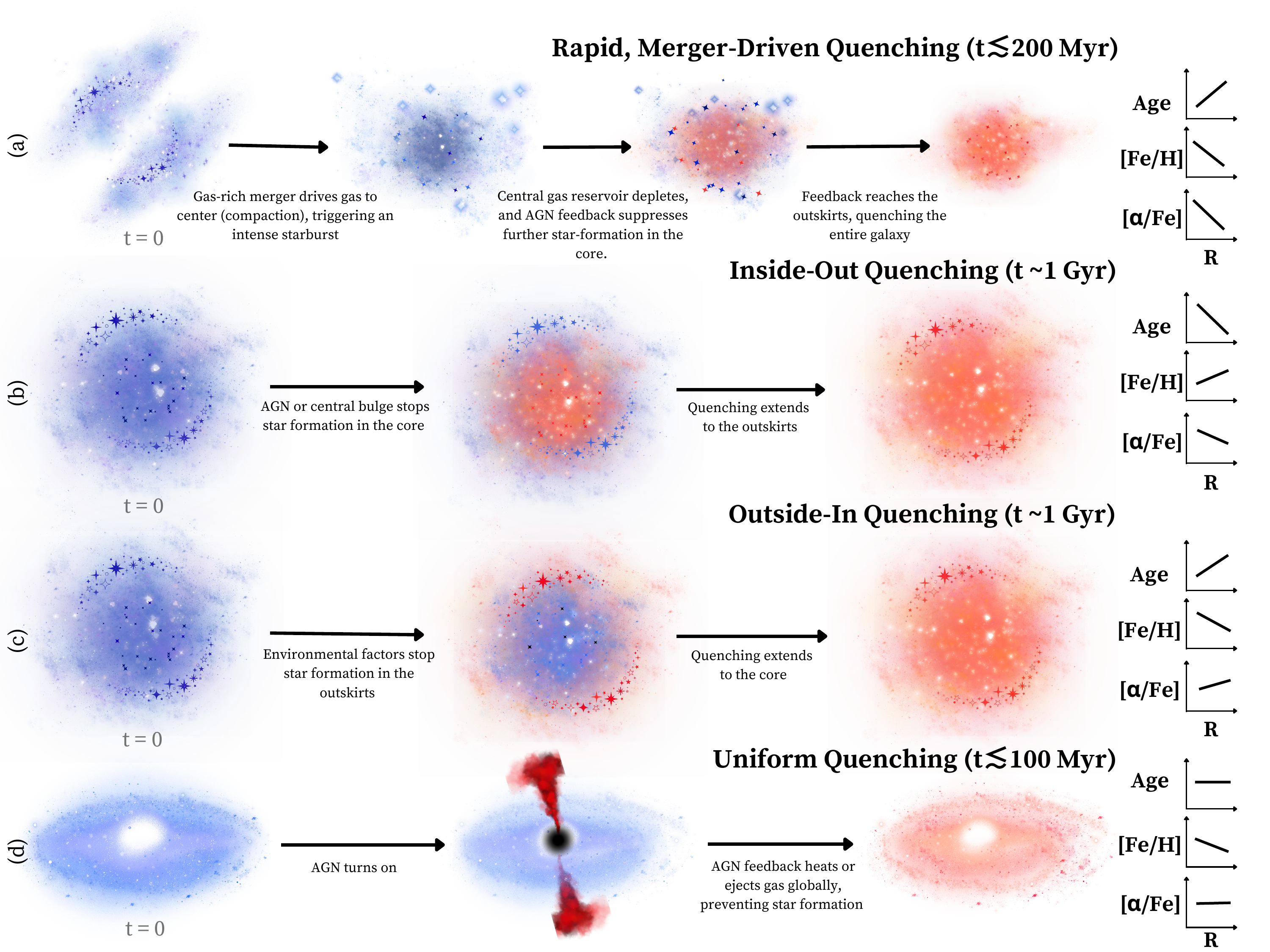}
\centering
\cprotect\caption{Schematic illustrating different quenching pathways possible for massive quiescent galaxies, along with the expected gradients and timescales involved.
\textit{Top row, (a)}: A rapid, merger driven quenching event. The expected gradients match those observed for XMM-VID1-2075 and XMM-VID3-1120.
\textit{Upper middle row, (b)}: An inside-out quenching event, where star formation in the central regions terminates first. 
\textit{Lower middle row, (c)}: An outside-in quenching event, where star formation in the outskirts ceases first. 
\textit{Bottom row, (d)}: A uniform quenching event, where an AGN quenches the host galaxy on very short timescales \citep[$\lesssim100$ Myr;][]{Setton20, DEugenio24}, producing gradients as observed in XMM-VID3-2457.
}
\label{fig:quenching_pathways}
\end{figure*}

Our results indicate that both uniform and rapid merger-driven quenching are already present in the first two billion years. This mirrors predictions from cosmological simulations for massive galaxies at $z \gtrsim 3.5$ \citep[i.e.,][]{Pathak21, Kimmig25, McClymont25, PAA25b} and observations at lower-$z$ \citep[i.e.,][]{Akhshik23}, suggesting that the diverse quenching pathways observed at low‑redshift were already established in the early Universe. We present a schematic to illustrate this in Figure \ref{fig:quenching_pathways} depicting four possible quenching pathways for MQGs, as well as how such pathways would produce gradients in age, [Fe/H] and [$\alpha$/Fe].

From a sample size of three MQGs, we see two distinct quenching pathways; rapid, merger-driven quenching, as well as uniform quenching. This seemingly indicates that a diversity of quenching mechanisms are already operating at early times. Larger sample sizes of MQGs using spatially resolved spectroscopy are needed to confirm this.

\subsubsection{Comparison to Lower-z Literature} \label{sec:lowerz_lit}

This work represents the first spatially resolved measurements of age, [Fe/H] and [$\alpha$/Fe] for massive quiescent galaxies at $z\gtrsim3$. To frame these results in the context of the larger picture of galaxy evolution, we compare our results to lower-$z$ literature.

At $z\sim0.6$, the SQuIGG$\overrightarrow{L}$E survey found that PSB galaxies exhibit flat $H\delta_{A}$ gradients \citep{Setton20}. As $H\delta_{A}$ is a proxy for stellar age, this implies that these PSBs quenched uniformly. Similar results have been shown at higher-$z$. 
For example, the REQUIEM-2D survey \citep{Akhshik20, Akhshik23} used spatially resolved HST spectroscopy to study 8 massive ($\rm log_{10}(M_{\ast}/M_{\odot}) > 10.6$) quiescent galaxies at $z\sim2$, finding that a subset of these MQGs have flat age gradients.
Likewise, \citet{DEugenio24} found no radial dependence in age in a massive ($\mathrm{log(M_{\ast}/M_{\odot} )\approx 11.2}$) PSB galaxy at $z\sim3$. 
These results are similar to what is found in XMM-VID3-2457, which is flat in age, and indicate that uniform quenching is common in PSBs.

In contrast, \citet{Cheng24} studied 456 massive ($10.3 \lesssim \mathrm{log(M_{\ast}/M_{\odot})} \lesssim 11.8$) quiescent galaxies at $0.6 \lesssim z \lesssim 1.0$ from the LEGA-C survey \citep{vdWel16, vdWel21, Straatman18}, measuring their age, [Fe/H], and [Mg/Fe] gradients. Over the full sample, they find flat age and [Mg/Fe] gradients, and negative [Fe/H] gradients. However, when splitting the sample into three equally sized age bins, the younger ($\rm 1 \leq age/Gyr < 2.63$) subsample has a slightly positive age gradient, and negative metallicity gradients, as shown in Figure \ref{fig:Radial_grad_all}. They use this to show that these young quiescent galaxies likely underwent a merger-triggered central starburst \citep{DEugenio20, Wu20}, similar to XMM-VID1-2075 and XMM-VID3-1120.

At $z\sim2$, \citet{Cheng25} measured age and abundance gradients for eight massive ($10.3 \lesssim \mathrm{log(M_{\ast}/M_{\odot})} \lesssim 11.1$) quiescent galaxies between $1.2 \lesssim z \lesssim 2.2$ using the JWST-SUSPENSE survey \citep{Slob24}. This work utilizes JWST/NIRSpec-MSA spectroscopy.
To obtain ages and abundances, they also use \texttt{alf}$\alpha$, but with the \citet{Conroy18} grids. 
They find their sample of quiescent galaxies exhibit negative age, positive [Mg/H] and [Mg/Fe], and flat [Fe/H] gradients, as depicted in Figure \ref{fig:Radial_grad_all}. They describe this as inside-out quenching, as the cores are older, and the low central [Mg/H] could indicate central gas expulsion \citep{Beverage21, Beverage25}. 

While this describes a similar quenching pattern to what is expected for our MQG sample, there is a clear difference in gradients observed between this work and \citet{Cheng25}. 
This is to be expected, as the galaxies studied in \citet{Cheng25} are much older ($\rm \langle age \rangle \approx 2$ Gyr) than the sample studied in this work ($\langle t_{\rm SSP} \rangle \approx 0.6$ Gyr). 
This in particular is key to understanding the significantly different gradients; the older sample would have had more time to be affected by events such as minor mergers. This could lead to the negative age gradients, where the inner-most regions are left relatively untouched, while the outskirts grow over time \citep{Haryana25}.

Furthermore, the MQGs in this work are already more massive ($\mathrm{log(M_{\ast}/M_{\odot})} > 11.2$) and more compact ($\rm R_e \lesssim2$ kpc) at a higher redshift than those in \citet{Cheng25}. We highlight this in Figure \ref{fig:mass-size}, where we plot the stellar mass and size reported in literature of massive quiescent galaxies. Galaxies of similar mass are almost all larger in spatial extent, while similarly compact galaxies are far less massive and are observed at later epochs. It follows then that our MQGs are not evolutionarily linked to the lower-$z$ populations discussed in \citet{Cheng25}, and should perhaps be expected to undergo different processes, resulting in different gradients. This interpretation is subject to progenitor bias, as the low-$z$ quiescent population includes larger and more recently quenched galaxies, which are not represented by this compact, high-$z$ MQG sample.

\section{Conclusions} \label{sec:conc}
In this work, we leverage JWST/NIRSpec IFU observations of three massive quiescent galaxies from the MAGAZ3NE survey at $z\sim3.5$ to measure age, [Fe/H] and [$\alpha$/Fe] gradients using \texttt{alf}$\alpha$ \citep{Beverage25, Beverage2024_AA} with the sMILES $\alpha-$enhanced models \citep{Knowles21, Knowles23}. We model their star-formation histories using \textsc{Bagpipes} \citep{Carnall18, Carnall19} using a flexible non-parametric Gaussian Process SFH \citep{Iyer19} and a smooth non-parametric continuity SFH \citep{Leja19}. Our main findings are summarized as follows:

\begin{enumerate}

    \item[(i)] All three galaxies are extremely massive ($\rm M_{\ast} > 10^{11.2}~\mathrm{M}_{\odot}$) with high velocity dispersions ($\sigma_\ast> 325$ km/s), in agreement with previous results using Keck/MOSFIRE spectroscopy \citep{Forrest20, Forrest22}  (Table \ref{tab:Bagpipes_Re_results}). These masses are in agreement with the most massive galaxy expected from the XMM-VIDEO survey area (Fig. \ref{fig:Re_SFH_MAH}). 
    They are young, with mass-weighted ages $t_{\rm MW} \lesssim 700$ Myr and recently quenched ($\Delta t_q = t_{obs} - t_q \approx 0.11-0.24$ Gyr).
    All three formed rapidly ($\langle t_{\rm form} \rangle = 422\pm34$ Myr) and quenched soon thereafter ($t_q - t_{\rm 50} \lesssim 500$ Myr).

    \item[(ii)]XMM-VID1-2075 ($[\rm \alpha/Fe] = +0.13^{+0.09}_{-0.08}$) and XMM-VID3-1120 ($\rm [\alpha/Fe] = +0.08^{+0.08}_{-0.20}$) are moderately $\alpha$-enhanced, while XMM-VID3-2457 is consistent with a solar-scaled abundance pattern ([$\alpha$/Fe] = $-0.02^{+0.05}_{-0.05}$) (Table \ref{tab:AA_Re_results}). 
    This is somewhat surprising, as massive quiescent galaxies observed at high-$z$ are commonly $\alpha$-enhanced \citep[${\rm [\alpha/Fe] \gtrsim +0.3}$; i.e.,][]{Kriek16,Beverage25,Kimmig25}. However, a spread in $\alpha$-enhancement has been observed \citep[$\langle \rm {[\alpha/Fe]} \rangle = + 0.22^{+0.22}_{-0.17}$;][]{Hamadouche26}, reflecting a diversity in evolutionary paths. This work adds to the body of literature that shows this diversity.

    \item[(iii)] The central regions of the three MQGs are remarkably similar, with young ($ \langle t_{\rm SSP} \rangle \approx 0.6$ Gyr) ages, elevated ($\gtrsim +0.21$) [$\alpha$/Fe], and solar metallicities ($\rm [Fe/H] \approx 0$) for two of the MQGs (Fig. \ref{fig:Radial_grad}). This is consistent with a recent, intense starburst. This is supported by the SFHs, which show a large burst ($\rm SFR_{peak} \approx 500 \, M_{\odot} \, yr^{-1}$) of star-formation between $4 < z < 5$ (Fig. \ref{fig:Re_SFH_MAH}). The radial SFHs agree with this picture, with central surface SFR densities that peak above $\Sigma_{\rm SFR} > 50 \, \rm M_{\odot} \, yr^{-1} \, kpc^{-2}$.
    This large burst is consistent with a previous sub-millimeter galaxy phase, which are commonly considered a plausible progenitor to MQGs. 
    
    \item[(iv)] We find that both XMM-VID1-2075 and XMM-VID3-1120 have statistically similar age and [Fe/H] gradients, while XMM-VID3-2457 exhibits statistically different age, [Fe/H] and [$\alpha$/Fe] gradients than the other two galaxies. In particular, both XMM-VID1-2075 and XMM-VID3-1120 exhibit positive age gradients and negative [$\alpha$/Fe] gradients (Fig. \ref{fig:Radial_grad}), consistent with a rapid, merger-driven quenching event (Fig. \ref{fig:quenching_pathways}).

    \item[(v)] XMM-VID3-2457 exhibits a flat age gradient (Fig. \ref{fig:Radial_grad}), suggestive of uniform quenching. Furthermore, its core is significantly $\alpha$-enhanced ($\rm[\alpha/Fe] = +0.52^{+0.06} _{-0.17}$) and Fe-deficient ($\rm[Fe/H] = -0.42^{+0.21} _{-0.14}$).
    This galaxy was quenched uniformly on very short ($\Delta t_q \lesssim100$ Myr) timescales (Figures \ref{fig:Radial_grad}, \ref{fig:UMG_SFH_MAH_radial}).
    Internal mechanisms are the most likely cause for quenching. The spectrum provides some evidence for an AGN, with \textsc{[O iii]} emission (EW$_{\rm rest}$ $= 3.6$ \AA) and \textsc{[N ii]} emission (EW$_{\rm rest}$ $= 3.5$ \AA).  

    \item[(vi)] A significant discrepancy ($\approx 0.2-0.4$ dex) appears between the reported metallicities from \textsc{Bagpipes} ($\langle \log_{10}(Z/Z_{\odot}) \rangle = 0.31$) and \texttt{alf}$\alpha$ ($\langle [\rm Fe/H] \rangle = 0.02$) (Tables \ref{tab:Bagpipes_Re_results} \& \ref{tab:AA_Re_results}). This is similar to other recent results showing that the metallicity measurements between codes using solar-scaled versus $\alpha$-enhanced models differ significantly \citep{Beverage25, McConachie25, Leung26b}. We note, however, that \textsc{Bagpipes} mass-weighted ages and \texttt{alf}$\alpha$ SSP-equivalent ages agree within $\approx1\sigma$. This offset in inferred metallicities impacts all analyses of the star formation histories of high-$z$ quiescent galaxies, and future work is required to further investigate this.

\end{enumerate}

This work presents the first spatially resolved analysis of stellar population gradients of massive quiescent galaxies in the first 2 Gyr of cosmic time. Despite our small sample size, these results highlight the importance of spatially resolved spectroscopy and $\alpha$-enhancement consideration in probing massive galaxy formation in the early Universe.


\begin{acknowledgments}
We thank Chloe Cheng for providing LEGA-C data. We also thank Kartheik Iyer for discussions on how to use the flexible non-parametric GP-SFH models. JAD acknowledges funding from the Dunlap Institute, made possible through an endowment established by the David Dunlap family and the University of Toronto. 
AM acknowledges support from the Yavin Family Fund.
BF acknowledges support from JWST-GO-02913.001-A.
DM acknowledges generous support from the Leonard and Jane Holmes Bernstein Professorship in Evolutionary Science. GW gratefully acknowledges support from the National Science Foundation through grant AST-2347348. 
\end{acknowledgments}




%
\facilities{JWST}

\software{\texttt{astropy} \citep{2022ApJ...935..167A}, \texttt{alf}$\alpha$ \citep{Beverage2024_AA, Beverage25}, \textsc{Bagpipes} \citep{Carnall18, Carnall19}, \texttt{statsmodels} \citep{statsmodels}
          }


\appendix

\section{Spectra} \label{sec:App_spec}

\begin{figure}[!htbp]
\includegraphics[width=\textwidth]{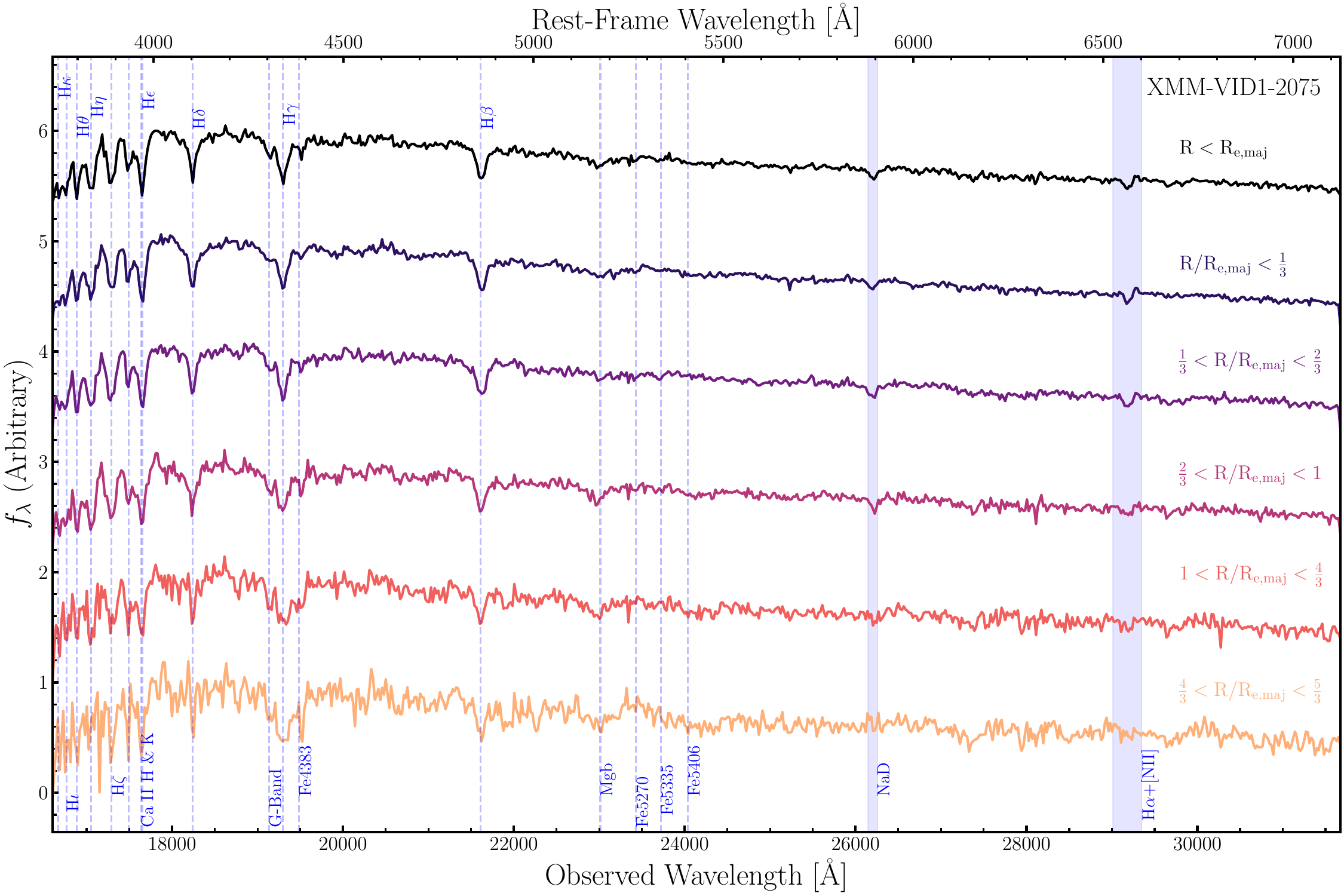}
\centering
\cprotect\caption{Extracted spectra for XMM-VID1-2075, within $R < R_{e}$ (\textit{top, black}) and elliptical radial bins, normalized and scaled arbitrarily for visual purposes. 
}
\label{fig:X2075_radial_spec}
\end{figure}

\begin{figure}[!htbp]
\includegraphics[width=\textwidth]{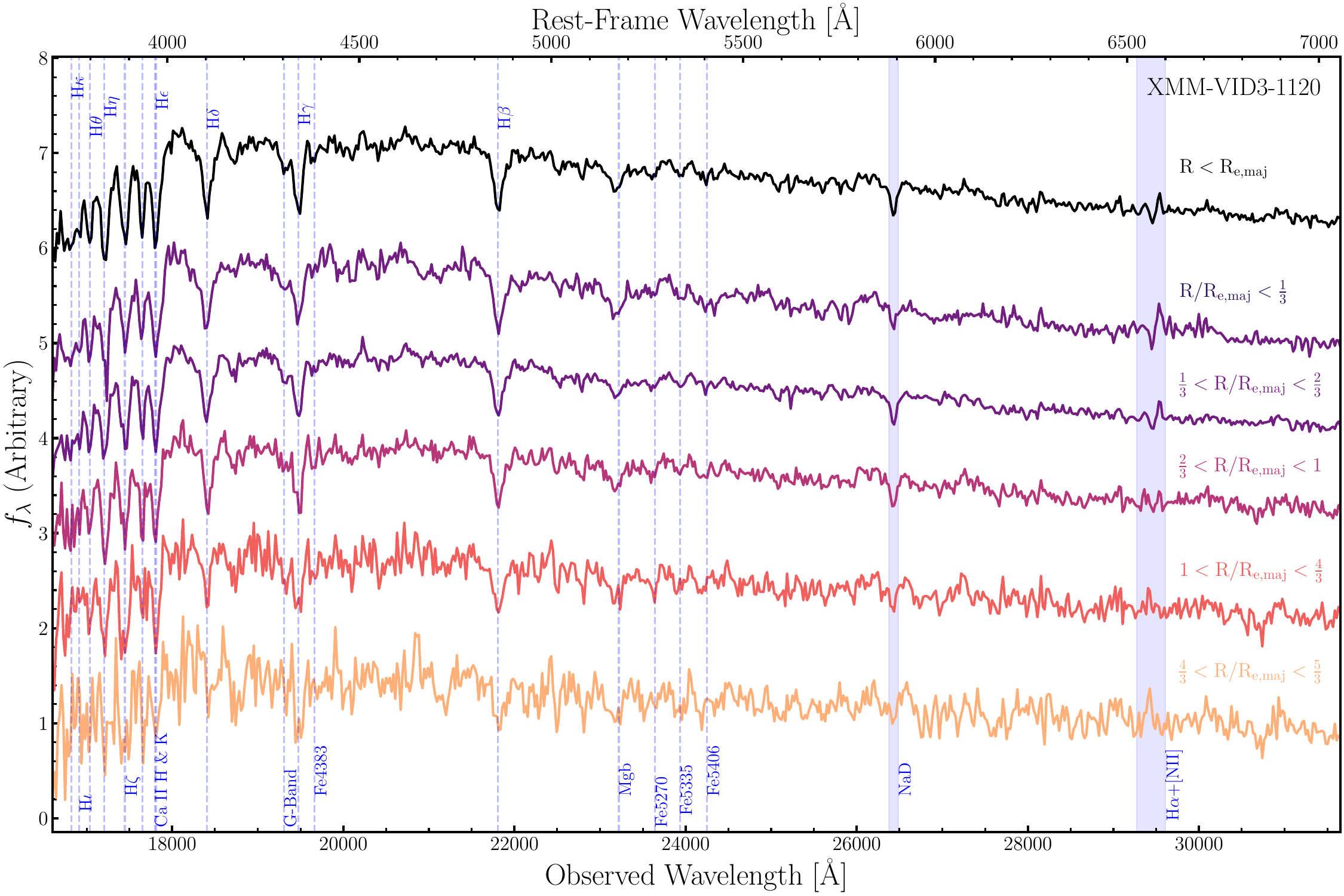}
\centering
\cprotect\caption{Same as Fig. \ref{fig:X2075_radial_spec}, but for XMM-VID3-1120.
}
\label{fig:X1120_radial_spec}
\end{figure}

\begin{figure}[!htbp]
\includegraphics[width=\textwidth]{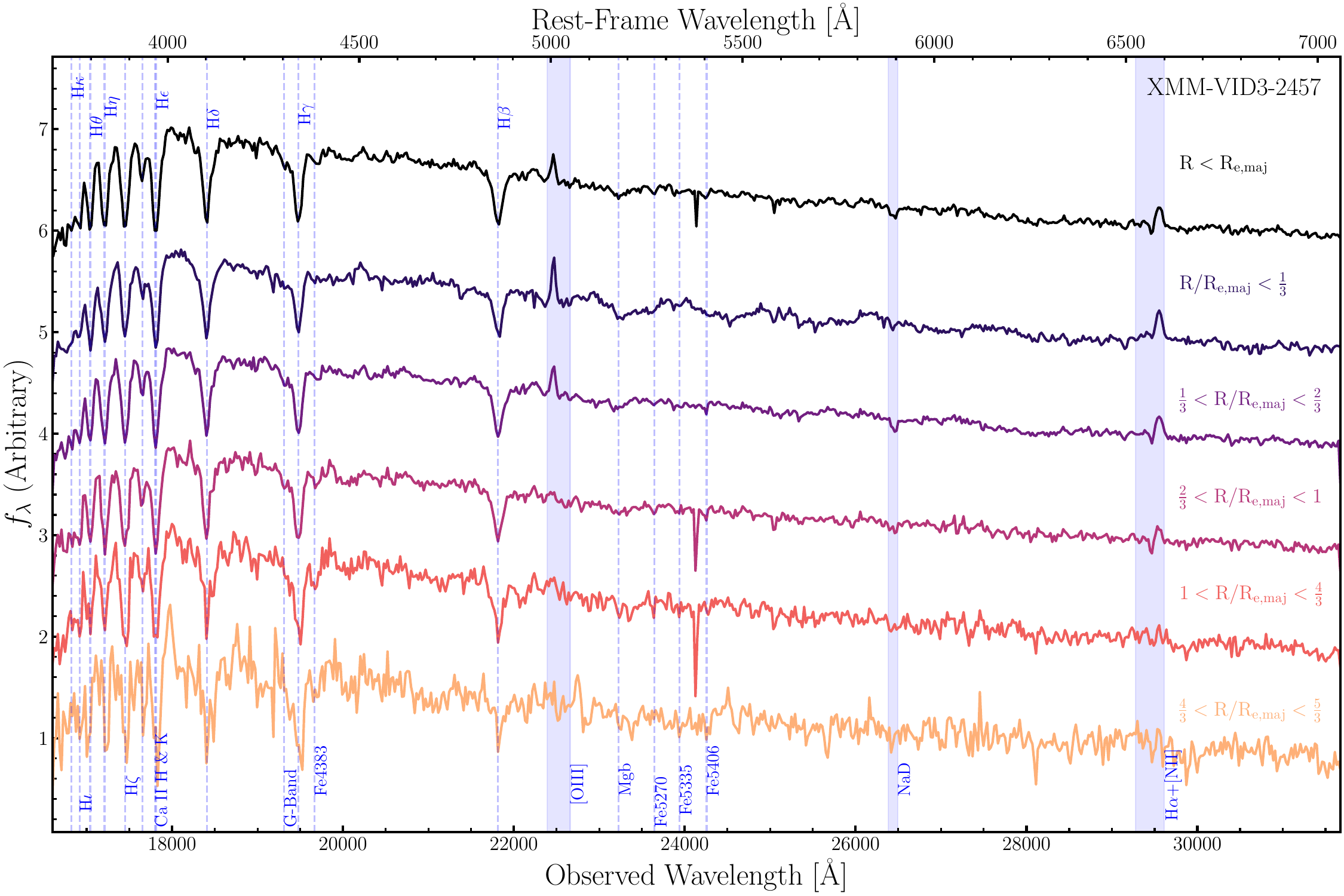}
\centering
\cprotect\caption{Same as Fig. \ref{fig:X2075_radial_spec}, but for XMM-VID3-2457.
}
\label{fig:X2457_radial_spec}
\end{figure}

\section{Tables} \label{sec:App_tables}

\subsection{Fitting  Priors} \label{sec:priors}

\begin{table*}[h!]
  \centering
  \begin{tabular}{@{} l l l l @{}} 
    \toprule
    \textbf{Component} & \textbf{Parameter} & \textbf{Range}                  & \textbf{Prior} \\ 
    \midrule
    \multicolumn{4}{l}{\textbf{Global}} \\
           & Redshift    & ($z_{\rm spec} - 0.1$, $z_{\rm spec} + 0.1$)                      & Uniform     \\
                 & Velocity Dispersion ($\sigma_{\rm \ast}$, km s$^{-1}$)     & (300, 600)                     & Gaussian ($\mu=400, \, \sigma=25$) \\
    \midrule
    \multicolumn{4}{l}{\textbf{Iyer et al.\ (2019) Non-parametric SFH}} \\
                 & SFR (M$_{\odot}$ yr$^{-1}$)         & (1\,$\times$\,10$^{-8}$, 2\,$\times$\,10$^4$) & $\log_{10}$ \\
                 & Bins        & 10                              & --          \\
                 & Bins\_prior & --                             & Dirichlet   \\
                 & t$\rm _{x}$          & [10, 20, 30, 40, 50, 60, 70, 80, 90]                   & --          \\
                 & $\alpha$       & [30, 3, 3, 3, 3, 3, 3, 3, 3, 3]                   & --          \\
                 & Stellar Mass ($\rm log(M_\ast / M_{\odot})$)  & (6, 13)                        & Uniform     \\
                 & Metallicity (Z$_{\odot}$) & (0.01, 5)                     &  $\rm log_{10}$     \\
    \midrule
    \multicolumn{4}{l}{\textbf{Continuity (Leja et al.\ 2019) Non-parametric SFH}} \\
                 & Bins        & 10                              & --          \\
                 & $\log_{10}(\rm SFR_{\it i}/SFR_{{\it i}+1})$  & $(-100, 100)$                     & Student-$t$ ($\sigma = 0.7$, $\nu=2$)   \\
                 & Stellar Mass ($\rm log(M_\ast / M_{\odot})$)  & (6, 13)                        & Uniform     \\
                 & Metallicity (Z$_{\odot}$) & (0.01, 5)                     &  $\rm log_{10}$     \\
    \midrule
    \multicolumn{4}{l}{\textbf{Dust}} \\
                 & Type        & Calzetti                              & --         \\
                 & Av          & (0, 1)                       & Uniform          \\
    \midrule
    \multicolumn{4}{l}{\textbf{Noise}} \\
                & Type          & Gaussian Process (SHOTerm)            &--          \\
                & White noise scaling ($s$)  & (1, 10)                  &log$_{10}$ \\
                & Correlated amplitude ($\sigma$) & (1\,$\times$\,10$^{-4}$, 1\,$\times$\,10$^{-1}$) & $\log_{10}$ \\
                & Period/length scale ($\rho$)  & (1\,$\times$\,10$^{-5}$, 1)   & log$_{10}$ \\
                & Dampening Quality Factor ($Q$) & 0.49 & Fixed \\
    \midrule
    \multicolumn{4}{l}{\textbf{Nebular}} \\
    & Ionization Parameter (log$_{10}$(U))          & $(-2, -4)$          & Uniform         \\
    \midrule
    \multicolumn{4}{l}{\textbf{AGN component\footnote{Only for XMM-VID3-2457.}}} \\
                 & $f_{\rm H\alpha, broad}\, (\rm erg\,s^{-1}cm^{-2})$                    & (0, 2.5\,$\times$\,10$^{-17}$) & Uniform     \\
                 & $f_{5100}\rm (erg\,s^{-1}cm^{-2}\mathring{A}^{-1})$                   & (0, 1\,$\times$\,10$^{-19}$)   & Uniform     \\
                 & $\sigma_{\rm AGN}$ (km s$^{-1}$)      & (1000, 7000)                   & $\log_{10}$ \\
                 & $\alpha_{\lambda<5000\rm\mathring{A}}$                 & $(-2, 4)$                        & Gaussian ($\mu=-1.5, \, \sigma=0.5$)    \\
                 & $\alpha_{\lambda>5000\rm\mathring{A}}$                  & $(-2, 2)$                        & Gaussian ($\mu=0.5, \, \sigma=0.5$)   \\
    \bottomrule
  \end{tabular}

  \footnotesize
  \caption{Priors used in \textsc{Bagpipes} spectral fitting.}
  \label{tab:bagpipes_priors}
\end{table*}

\begin{table}[ht]
\centering
\begin{tabular}{l l}
\hline
Parameter               & Prior Range                                   \\
\hline
Age (Gyr)               & $(10^{-1.5},\,\mathrm{t_{H}}(z))$ \\
$\sigma_{\ast}$ (km s$^{-1}$)  & $(100,500)$                                    \\
Recessional Velocity (km/s)    & $(-500,\,500)$                                 \\
Metallicity $[Z/\mathrm{H}]$ & $(-1.5,\,0.26)$                            \\
$[\alpha/\mathrm{Fe}]$   & $(-0.2,\,+0.6)$                                 \\
Jitter                  & $(0.1,\,10)$                                   \\
\hline
\end{tabular}
\caption{Prior ranges for the \texttt{alf$\alpha$} fits. The upper bound of the age parameter is determined by calculating for the age of the Universe at the redshift of the galaxy.}
\label{tab:alfalpha_priors}
\end{table}

\subsection{Fitting Results} \label{sec:Fitting_results}

\section{\texorpdfstring{alf$\alpha$ Testing}{alfa Testing}} \label{sec:AA Testing} 

We compare the sMILES and MILES+IRTF \citep{Conroy18} grids in order to understand the systemic differences between these two stellar population models. We do this by creating a mock galaxy spectrum using \textsc{Bagpipes}, and then running it through \verb|alf|$\alpha$ with both grids. This allows us to set the age, metallicity and [$\alpha$/Fe] as a known prior, therefore allowing us to directly compare our ability to recover these parameters using each of these models. We create a mock galaxy spectrum at $z=2$, with age ($t_{\rm age} = 1.1$ Gyr), solar metallicity ($\rm Z_{\ast}= Z_{\odot}$), and velocity dispersion ($\sigma_\ast = 200$ km/s). \textsc{Bagpipes} uses BC03, which is solar-scaled,  [$\alpha$/Fe] = 0 by default. All specified values are within the constraints of both the sMILES and \citet{Conroy18} grids. We use  an exponential star formation history with $\tau =0.05$ Gyr. This is chosen to resemble how \texttt{alf}$\alpha$ models the SFH, which assumes a single burst. Finally, we add noise to the spectrum, resulting in a signal-to-noise $\approx 20$ \AA$^{-1}$ spectrum.

\begin{figure}[!htb]
\includegraphics[width=\textwidth]{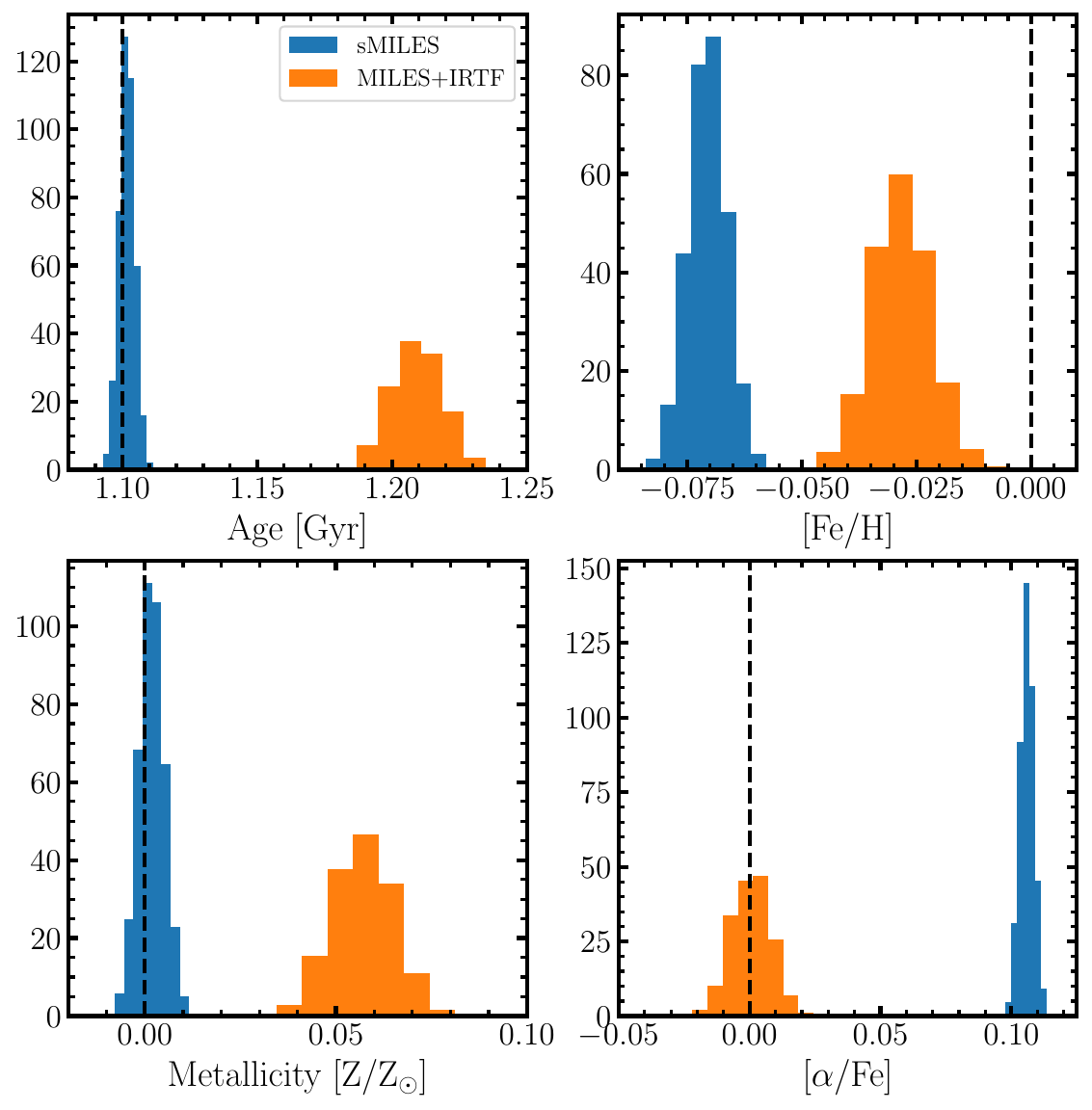}
\centering
\cprotect\caption{Results from fitting a model galaxy made in \textsc{Bagpipes} with \verb|alf|$\alpha$, using the MILES+IRTF models from \citet{Conroy18} (orange) and sMILES \citep{Knowles23} (blue) for age (top left), [Fe/H] (top right), metallicity (bottom left), and [$\alpha$/Fe] (bottom right). Posterior fitting distributions are presented as probability density functions. 
The model galaxy is made with an age of 1.1 Gyr, $\rm Z_{\ast}=Z_{\odot}$, [$\alpha$/Fe] = 0, and $\sigma_\ast = 200$ km/s.  
These values are indicated by a black dashed line.}
\label{fig:sMILESvConroy_results}
\end{figure}

We show the results of this testing in Fig. \ref{fig:sMILESvConroy_results} by plotting the posterior distributions of age ($t_{\rm age}$), [Fe/H], [Z/H], and [$\alpha$/Fe] for both the sMILES models as well as the MILES+IRTF models from \citet{Conroy18} as 1-D histograms. The input parameter values for the mock galaxy are shown as a dashed black line. sMILES better recovers total metallicity and age (top and bottom left panels, respectively) than the MILES+IRTF models, capturing the input parameters within 1$\sigma$. 
However, this is not true for [Fe/H] and [$\alpha$/Fe]. Although the input spectrum assumes a solar metallicity ($\rm [Fe/H]=0$), neither sMILES nor the MILES+IRTF model recovers this, returning values of [Fe/H] $ = -0.071^{+0.004}_{-0.004}$ and [Fe/H] $= -0.028^{+0.007}_{-0.006}$ respectively. 
Furthermore, sMILES struggles with recovering [$\alpha$/Fe], returning $[\alpha/\rm Fe] = +0.106^{+0.003}_{-0.003}$, while MILES+IRTF returns  $[\alpha/\rm Fe] = +0.001^{+0.007}_{-0.008}$, consistent with the input value of 0 from Bagpipes. This offset of $\approx 0.1$ dex in both [Fe/H] and [$\alpha$/Fe] reflects important differences between the underlying stellar population models.

\begin{figure}[!htbp]
\includegraphics[width=\textwidth]{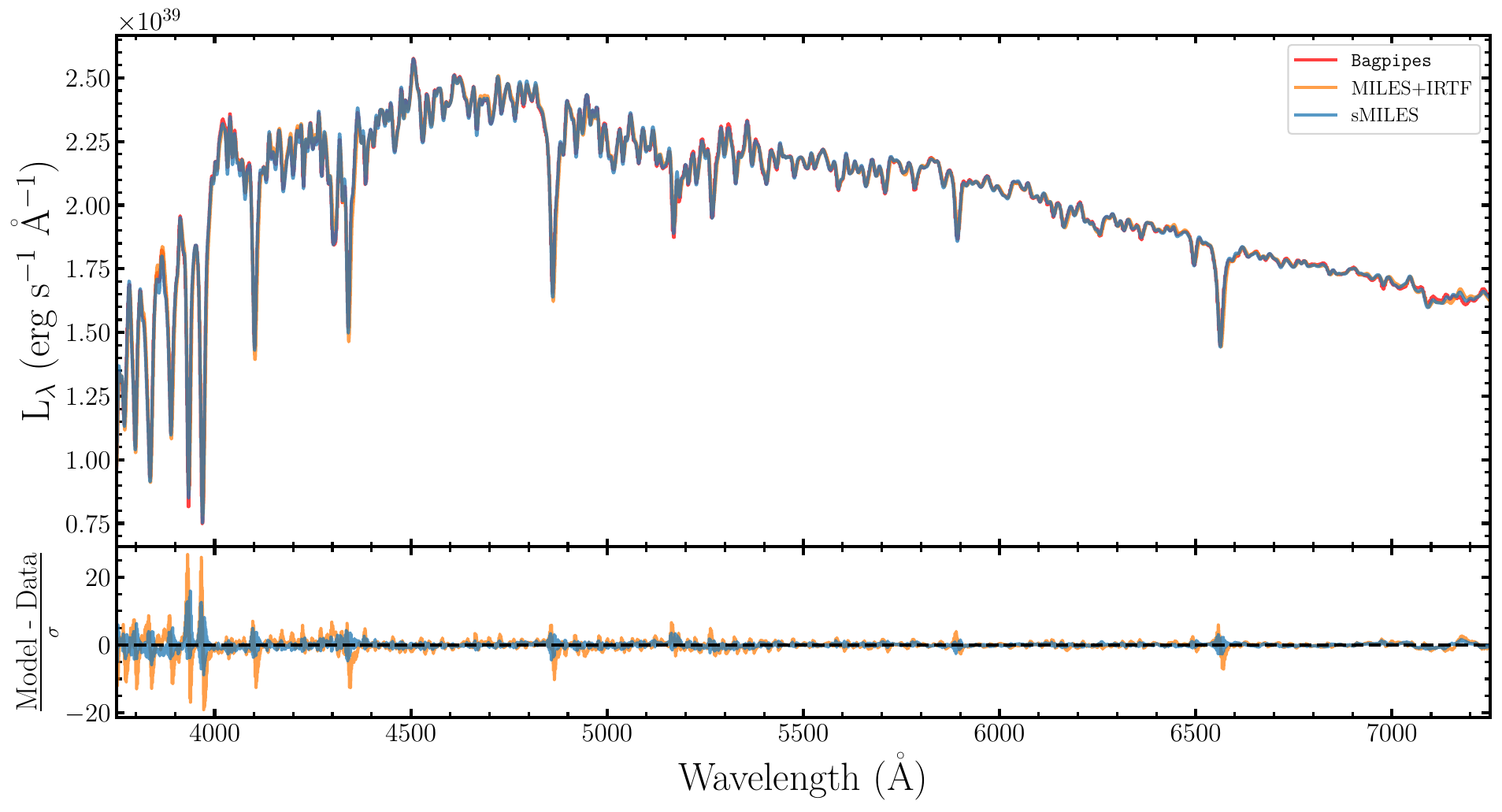}
\centering
\cprotect\caption{\textit{Top}: Red: \textsc{Bagpipes} model spectrum of a 1.1 Gyr galaxy with $\rm Z_{\ast}= Z_{\odot}$, [$\alpha$/Fe] = 0, and $\sigma_{\ast}$ = 200 km/s. MILES+IRTF (orange) and sMILES (blue) fits from \verb|alf|$\alpha$. 
\textit{Bottom}: The residuals between the MILES+IRTF and sMILES fits and the model galaxy. Significant differences between the models are seen, particularly for Balmer absorption line depths. The MILES+IRTF spectrum has deeper Balmer lines than either sMILES or \textsc{Bagpipes}.  
}
\label{fig:Grid_spec_comp}
\end{figure}

To understand the cause of this offset, we create model spectra using the MILES+IRTF and sMILES grids that have similar parameters as the \textsc{Bagpipes} mock galaxy spectrum. 
We show these model spectra in Figure \ref{fig:Grid_spec_comp}. The spectra of all three models appears largely similar. However, they differ non-trivially, particularly at Balmer absorption line depths. This should be expected, as the model spectra are made from SSPs that use different isochrones and stellar libraries. We discuss this more in Section \ref{sec:Isochrones+StellarLibs}.


\subsection{Effects of different Isochrones \& Stellar Libraries} \label{sec:Isochrones+StellarLibs}

Two stellar population synthesis (SPS) codes applied to the same spectrum can reasonably return different parameters due to differences in the isochrones and stellar libraries. 
Previous literature has studied such systematic offsets when observing local galaxies. \citet{Asad20} found that for galaxies aged $\sim$ 2-5 Gyr, the age measured using MIST \citep{MISTv1_2} isochrones is always +0.2 dex above the age measured using Padova \citep{PadovaIsoI, PadovaIsoII} isochrones even when using the wavelength range of 3700 $\text{\AA}$ to 5000 $\text{\AA}$. \citet{Palakkatharappil23} shows a similar offset between measured values resulting from different isochrones for the same spectra of stars in the NGC 2477 open cluster. They found that BaSTI \citep{BaSTI_I_solar, BaSTI_II_alpha} isochrones prefer an age $\approx$ 0.2 dex younger than MIST isochrones, as well as a metallicity that is $\sim$ 0.1 dex higher. 
\citet{Knowles23} also found a similar offset. They compared the sMILES grids to that of \citet{Conroy18} and MILES \citep{Vazdekis15} by taking the ratio of $\alpha$-enhanced and solar-scaled SSP spectra (with solar metallicity and an age of 9 Gyr) for all three grids. They found that all three grids return ratios that exhibit similar trends with increasing [$\alpha$/Fe]. However, \citet{Knowles23} notes that there are differences between the three. For example, sMILES and \citet{Conroy18} models predict the EW of H$\beta$ increases with increasing [$\alpha$/Fe], while the MILES models show decreasing H$\beta$. They note that these differences require further study. 

We are just starting to understand the impact of these systematic differences at high-$z$, with this work adding to the body of literature investigating this. Moreover, beyond just the inclusion of different stellar libraries and isochrones, consideration for how $\alpha$ elements are treated (i.e., varied individually or in lock-step) is important. 
For example, a comprehensive study of different stellar population libraries is presented in \citet{Jafariyazani25}. Using a sample of $z\sim2$ quiescent galaxies, they find that, on average, the \citet{Conroy18} models report [Mg/Fe] values $\approx 0.3$ dex above the [$\alpha$/Fe] from MILES \citep{Vazdekis15} models. They also found that the \citet{Conroy18} models report [Fe/H] values $\approx 0.3$ dex lower than what is reported by MILES. They explain the offset between [Mg/Fe] and [$\alpha$/Fe] as being due to the MILES models varying $\alpha$-element abundances in lock-step, similar to what is done by sMILES \citep{Knowles21, Knowles23}. When accounting for this, they find offsets of only $\sim0.1$ dex in most cases between the MILES and \citet{Conroy18} models for [$\alpha$/Fe]. However, the offsets in [Fe/H] are not explained by this, and \citet{Jafariyazani25} notes that this requires further investigation into systemic model uncertainties.

\begin{figure}
    \centering
    \includegraphics[width=\textwidth]{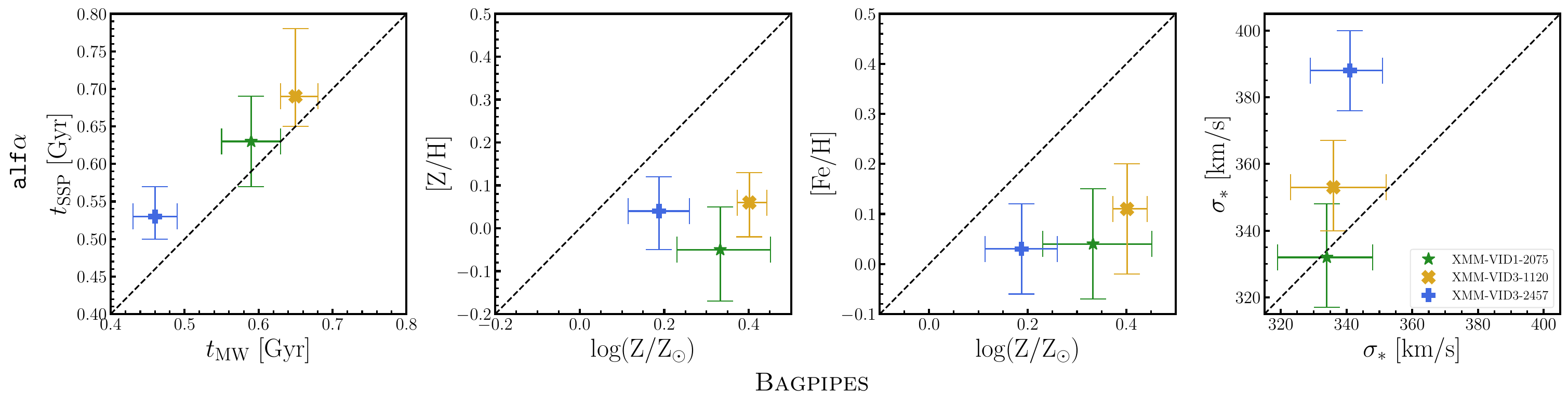}
    \caption{A comparison of different stellar population parameters inferred from  \texttt{alf}$\alpha$ (y-axis) and \textsc{Bagpipes} (x-axis) for our integrated spectra ($\rm R < R_e$). The one-to-one line is indicated as a black dashed line. 
    \textit{Left}: SSP-equivalent ages from \texttt{alf}$\alpha$ compared to the mass-weighted ages from \textsc{Bagpipes}. 
    \textit{Center-left}: [Z/H] inferred from \texttt{alf}$\alpha$ versus log($\rm Z_{\ast}/Z_{\odot}$) from \textsc{Bagpipes}. 
    \textit{Center-right}: [Fe/H] versus log($\rm Z_{\ast}/Z_{\odot}$) for the three galaxies. 
    \textit{Right}: A comparison of velocity dispersions $\sigma_{\ast}$ measured from \texttt{alf}$\alpha$ and \textsc{Bagpipes}. Both take into account the wavelength dependent nature of the JWST/NIRSpec IFU instrumental dispersion.
    }
    \label{fig:alfavsbagpipes}
\end{figure}

We conclude that this difference in measured values based on the choice of isochrones and stellar libraries is the cause of the $\approx$ 0.1 dex offset in metallicities and [$\alpha$/Fe] between the two grids. This unavoidable systematic affects all analyses of the ages and star formation histories of high-$z$ quiescent galaxies.
However, this offset does not explain the discrepancies found in this work and others \citep[i.e.,][]{Beverage25, McConachie25b, Leung26} between metallicities inferred from codes that use solar-scaled stellar libraries and those that allow variable elemental abundances (See \S \ref{sec:metals}). These offsets can be as large as $\sim0.9$ dex \citep{Beverage25, Leung26}. While the offsets between \textsc{Bagpipes} and \texttt{alf}$\alpha$ found in this work are not as severe, they do range from $\approx 0.2-0.4$ dex. We illustrate the difference of inferred stellar populations in Fig. \ref{fig:alfavsbagpipes}.
The ages returned by both codes agree well, with the \texttt{alf}$\alpha$ SSP-equivalent ages slightly higher, though consistent within $1\sigma$. 
\textsc{Bagpipes} metallicities are higher than \texttt{alf}$\alpha$ [Fe/H] and [Z/H] for all three MQGs, with offsets of $\approx 0.2-0.4$ dex. 
$\sigma_{\ast}$ is consistent within 1$\sigma$ between the two codes for two of the three MQGs.

\bibliography{BobbyBib}{}
\bibliographystyle{aasjournalv7}



\end{document}